\documentclass[twocolumn]{aastex631}

\usepackage[T1,T2A]{fontenc}
\usepackage{graphicx}
\received{}
\revised{}
\accepted{}
\submitjournal{ApJ}

\shorttitle{Near IR and X-ray Variability of Cyg~X-3.}
\shortauthors{Antokhin at al.}

\begin{document}

\title{Near-IR and X-ray Variability of Cyg~X-3: Evidence for a Compact IR Source and Complex Wind Structures}

\author[0000-0002-3561-8148]{Igor I. Antokhin}
\affil{Moscow Lomonosov State University, \\
Sternberg State Astronomical Institute, \\
119992 Universitetskij prospect, 13, Moscow, Russian Federation}

\author[0000-0001-5595-2285]{Anatol M.Cherepashchuk}
\affil{Moscow Lomonosov State University, \\
Sternberg State Astronomical Institute, \\
119992 Universitetskij prospect, 13, Moscow, Russian Federation}

\author[0000-0002-7150-8640]{Eleonora A. Antokhina}
\affil{Moscow Lomonosov State University, \\
Sternberg State Astronomical Institute, \\
119992 Universitetskij prospect, 13, Moscow, Russian Federation}

\author[0000-0002-4398-6258]{Andrey M. Tatarnikov}
\affil{Moscow Lomonosov State University, \\
Sternberg State Astronomical Institute, \\
119992 Universitetskij prospect, 13, Moscow, Russian Federation}

\correspondingauthor{Igor I. Antokhin}
\email{igor@sai.msu.ru}

\begin{abstract}

We study near-infrared ({\em JHK}) and X-ray light curves of Cyg X-3 obtained with the 2.5-m telescope of the Caucasian Mountain Observatory of MSU SAI and collected from {\em RXTE} ASM and {\em MAXI} archives. The light curves in the X-ray and IR domains are strongly affected by irregular variations. However, the mean curves are remarkably stable and qualitatively similar in both domains. This means that the IR flux of the system originates not only from the free-free radiation of the WR wind but also from a compact IR source located near the relativistic companion. The shape of the mean X-ray and IR light curves suggest the existence of two additional structures in the WR wind -- a bow shock near the relativistic companion and a so-called ``clumpy trail'' \citep{vilhu13}. Modeling of the mean X-ray and IR light curves allowed us to obtain important system parameters: the orbital phase of the superior conjunction of the relativistic companion $\phi_0=-0.066\pm 0.006$, the orbital inclination angle $i=29.5^\circ\pm 1.2^\circ$, and the WR mass-loss rate \mbox{$\dot{M} = (0.96\pm 0.14)\times 10^{-5}\,\rm M_\odot yr^{-1}$}. By using relations between $\dot{M}$ and the rate of the period change and between $\dot{M}$ and the WR mass, we estimated the probable mass of the relativistic companion $M_{\rm C}\simeq 7.2\,\rm M_\odot$ which points towards the black hole hypothesis. However, this estimate is based on the assumption of a smooth WR wind. Considering the uncertainty associated with clumping, the mass-loss rate can be lower which leaves room for the neutron star hypothesis.

\end{abstract}


\section{Introduction} \label{sec:intro}

Cyg~X-3 (WN4-8+c, period $P\simeq 4.8$~hr) located at a distance $\sim 7$~kpc \citep{ling09} is the only known X-ray binary in our Galaxy which components are a WR star and a relativistic companion (either a neutron star or a black hole). To date, two more massive X-ray binaries have been discovered in other galaxies containing WR stars as a donor star: \mbox{IC 10 X-1} (WNE+c, $P = 1.5\,\rm days$, \citealt{carpano19}) and \mbox{NGC 300 X-1} (WN5+c, $P = 1.3\,\rm days$, \citealt{steiner16}). The system was discovered in 1966 \citep{giacconi67} but for many years the type of the optical companion remained unknown due to extremely large interstellar absorption which made it impossible to observe the system in the optical domain. For one of the first reviews of the studies devoted to Cyg X-3, see \cite{bonnet88}.

Cyg X-3 has been a subject of many studies from radio to gamma domains. The orbital modulation was found in X-ray Uhuru data by \cite{pars72}. Shortly after, a strongly variable radio source was discovered near the position of Cyg X-3 \citep{baes72}. The orbital modulation in near infrared ($2.2\,\rm \mu m$) was discovered by \cite{becklin73, becklin74} who also identified the radio source with the system. The relativistic double-sided jet was discovered in the radio by \cite{molnar88} and further investigated by \cite{schal95}, \cite{mio01}, \cite{miller04}, \cite{tudose07} and others. Orbital variability in gamma-rays was discovered by \cite{fermi09} who also found the correlation of the gamma-ray flux with the radio emission from the relativistic jets. For more details and references see e.g. \cite{kol17}, \cite{vilhu09}, \cite{zd18}. The variability of the source is complicated by strong irregular changes of the flux and by state changes in the X-ray and gamma domains. However, as was first shown by \cite{klis81}, the mean X-ray light curve averaged over many orbital cycles is remarkably stable. This means that the underlying machinery related to the orbital motion of the components and accretion onto the relativistic companion, when considered on long time scales, remains the same. The X-ray and IR light curves exhibit one minimum during an orbital cycle, corresponding to the position of the compact object behind the WR star. No secondary minimum is observed implying that the orbital inclination angle is probably less than $\sim 70^\circ$.

The companion was identified as a WR star by \cite{vankerk93, vankerk96} thanks to IR spectroscopy. Given that the X-ray luminosity of the compact object can be as high as $L_x=5\times 10^{38}$~erg\,s$^{-1}$ \citep{vilhu09}, the part of the WR wind illuminated by the X-ray source should be highly ionized. Indeed, as shown by \cite{vankerk93, vankerk96}, the IR spectrum of the system drastically changes with the orbital phase, showing ``normal'' WR spectral features only at those orbital phases where the part of the WR wind shadowed from the X-ray radiation by the body of the star, is in front. At other phases the IR spectrum of the binary is strongly variable and indicates much higher state of ionization.

Following these findings, \cite{zd12} explained the X-ray minimum by the variable absorption of the X-ray emission in the (spherically symmetric) WR wind. In the IR domain, the minimum was explained by variable free-free emission of the two component WR wind (hot part illuminated by the X-ray source and cold part shadowed by the WR body) by \cite{vankerk93, vankerk96}.

As the amount of X-ray data increased, the mean X-ray light curve became more and more accurate. In particular, one more small minimum was discovered on the X-ray light curve around the orbital phase 0.4. It was explained by \cite{vilhu13} as being due to absorption of the X-ray flux of the relativistic companion by dense wind clumps formed by jet bow shocks (the so-called ``clumpy trail''). Another interesting feature of the mean X-ray light curve is that the ingress of the primary minimum is faster than the egress. This cannot be explained by absorption of X-rays in the spherically symmetric WR wind as the latter produces a symmetric light curve \citep{zd12}.

Most relatively recent studies in the IR domain were spectroscopic with only a few dealing with IR light curves. \cite{vankerk93, vankerk96} composed their IR light curves from the line-free areas of their flux calibrated IR spectra. \cite{matz96} obtained {\em JHK} photometry over nearly one single orbital cycle.  \cite{fender96} obtained IR photometry in several bands in a single observation covering the orbital phases $0.7-0.8$. \cite{koch02} obtained a single $2.4-12\,\rm\mu m$ light curve in the phase interval $0.83-1.04$ from their spectrophotometric observations.

The purpose of the present study was to obtain several IR light curves of the system in order to look at their variability from one orbital cycle to another, in order to determine how stable the mean light curve is and to compare its shape with the mean X-ray light curve. The next goal was to find out if we can get useful information about the system and its parameters from the analysis of the mean X-ray and IR light curves. 

The paper is organized as follows. In Section~\ref{sec:obs} we describe the observations used in the analysis. In Section~\ref{sec:meanlc} we compose and present the mean X-ray and IR light curves. In Section~\ref{sec:model} we first discuss the potential model qualitatively and then present the model used to fit the data. In Section~\ref{sec:results} the results of the model fitting are presented. The caveats of the model are discussed in Section~\ref{sec:caveats}. We proceed with the discussion of our results in Section~\ref{sec:discussion}. Section~\ref{sec:concl} presents a summary of our study.

\section{IR Observations and X-ray Archival Data}\label{sec:obs}

Infrared photometry of Cyg X-3 was carried out in 2016 and 2017 at the 2.5-m telescope of the Caucasian Mountain Observatory (CMO) of the Lomonosov Moscow State University Sternberg State Astronomical Institute (MSU SAI). The telescope is equipped with an infrared camera-spectrograph ASTRONIRCAM \citep{nad17}. It was used in a photometric mode with {\em JHK} bands of the Mauna Kea Observatories photometric system. A total of 14 nights of observations were obtained. In most cases we recorded a complete orbital cycle ($\sim$4.8~hr) overnight (except a few cases owing to weather conditions). Also, in order to record a potential rapid variability, during 11 nights out of 14, only {\em JK} (8 nights) or {\em J} (3 nights) bands were used. 

The usual data reduction procedure (bias subtraction, flat field correction) was performed, followed by aperture photometry. The differential magnitudes were calculated relative to the primary 2MASS comparison star c1 = J20322359+4057156 ($J_{MKO}=12.283$, $H_{MKO}=11.700$, $K_{MKO}=11.495$). {\em JHK} magnitudes of c1 in the $MKO$ photometric system where obtained by transformation from the 2MASS magnitudes \citep{leggett06} combined with additional corrections  by using the nearby bright stars J20322573+4058063, J20322766+4057111, and J20322807+4057424. To control the absence of c1 variability, two more control stars from the same field of view were used with colors comparable to those of Cyg~X-3.

Signal-to-noise ratio of measured count rates for Cyg X-3 and comparison stars was higher than 200 in all bands. However, the main source of uncertainty in differential magnitudes was inaccuracy of the flat field. First, in the {\em K} band, the flat field depends on the temperature of various parts of the telescope and its mount, the position of which changes during the night. For this reason, we were forced to calculate flat fields from frames with stars, filtering out the latter. Second, the HgCdTe detector of the camera suffers from an afterglow effect. To minimize it and facilitate the construction of flat fields, each exposure consisted of several short exposures with 5\arcsec ~ shifts between them. Nevertheless, the afterglow also introduces some error in the resulting flat field. We did not perform reduction for the atmospheric extinction. The relatively narrow {\em JHK} bands, low atmospheric extinction in the near-IR, the presence of comparison and control stars within the field of view and the absence of strong atmospheric absorption features within the bands lead to very small atmospheric corrections (in our case, much less than 0.01 mag). The final measurement errors were estimated by the scatter of the control stars magnitudes and are equal to $\sim 0.02$, $\sim 0.02$, and $\sim 0.01$\,mag in {\em JHK} bands respectively.

The observation log is presented in Table~\ref{table1}. The last column of the table shows the number of individual measurements during a night in every IR band. The total number of individual measurements is equal to 2790, 413, 1709 in {\em JHK} bands respectively. The measured differential magnitudes (var-c1) are presented in Tables~\ref{table2}, \ref{table3}, \ref{table4}. The moments of observations were reduced to the solar system barycenter.

\begin{figure*}
\centering
\plotone{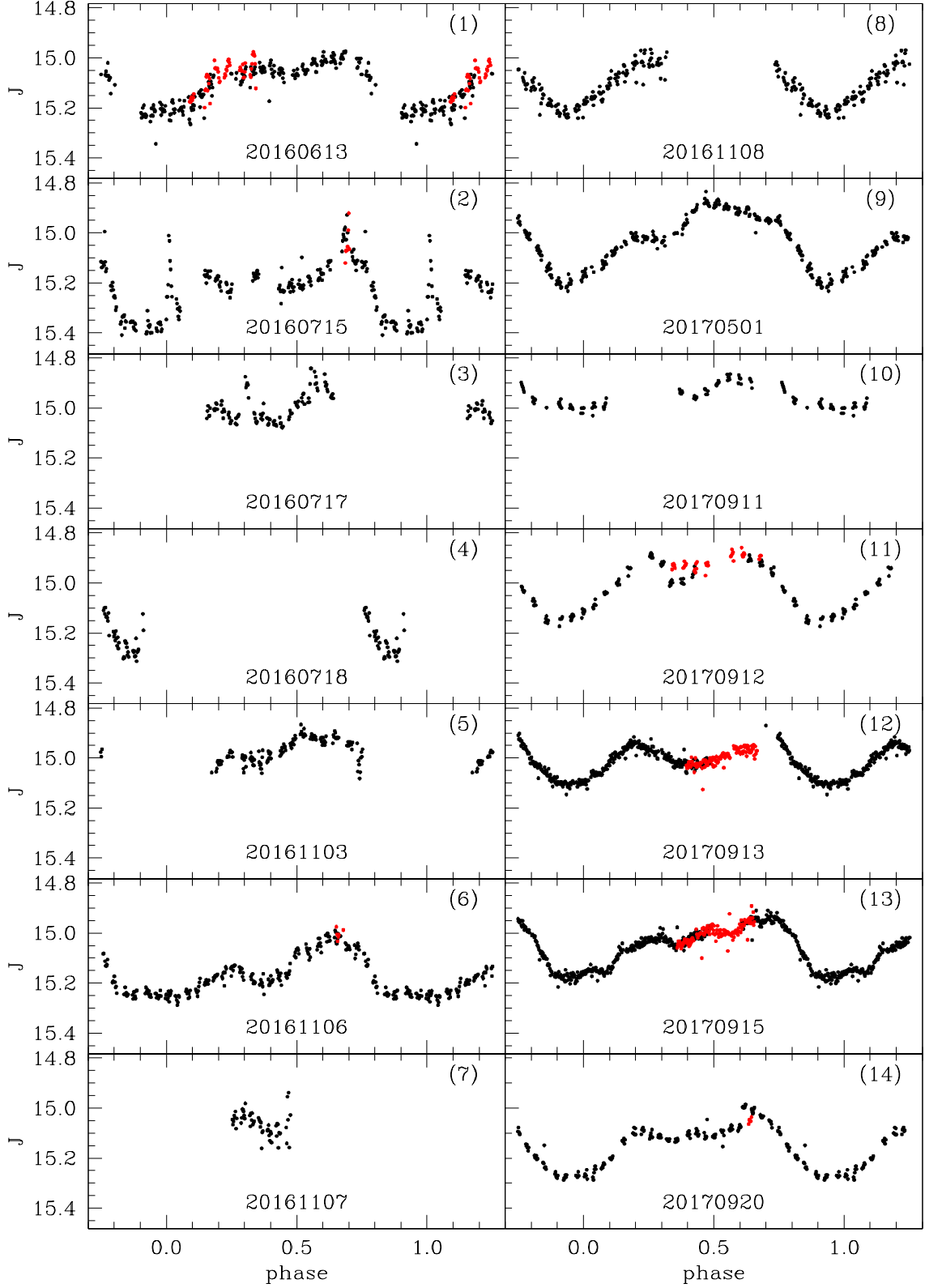}
\caption{Folded {\em J} light curves of Cyg~X-3 for all nights of observations. The observation dates are in the format YYYYMMDD. The magnitudes are obtained by adding the {\em J} c1 magnitude from Section~\ref{sec:obs} to the differential magnitudes. Note that the magnitude limits in all plots are identical; variations of the mean magnitude and light curve amplitude are evident. Data points that are further than one orbital period from the start of observations are shown in red (the same applies to Figs.~\ref{lc_jk_panel},~\ref{lc_jhk_panel}).}
\label{lc_j_panel}
\end{figure*}

\begin{figure*}
\centering
\plotone{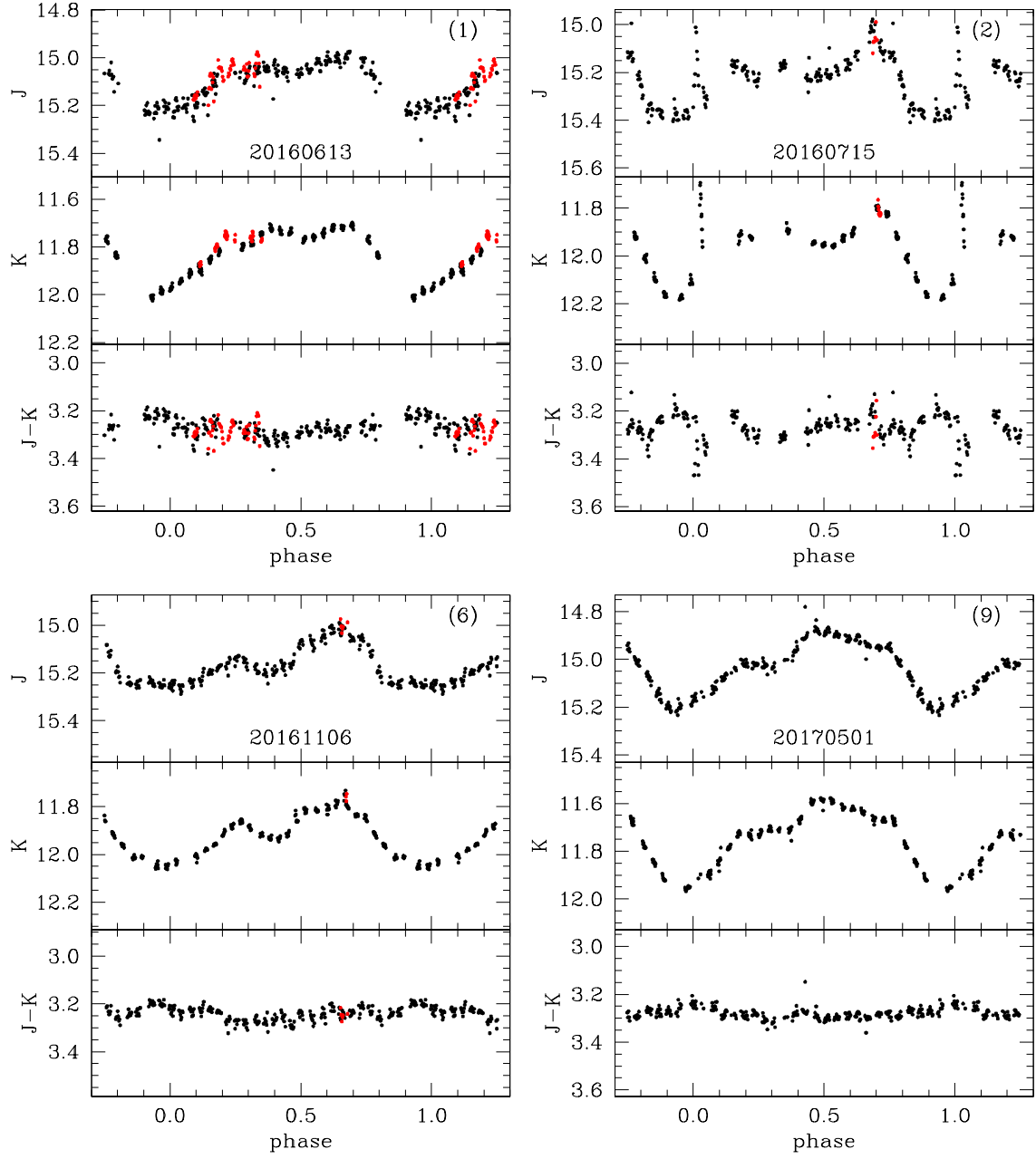}
\caption{Folded {\em JK} light curves of Cyg X-3 for for nights with data covering the entire orbital cycle. The numbering of the panels corresponds to that in Fig.~\ref{lc_j_panel}.}
\label{lc_jk_panel}
\end{figure*}

\begin{figure*}
\centering
\plotone{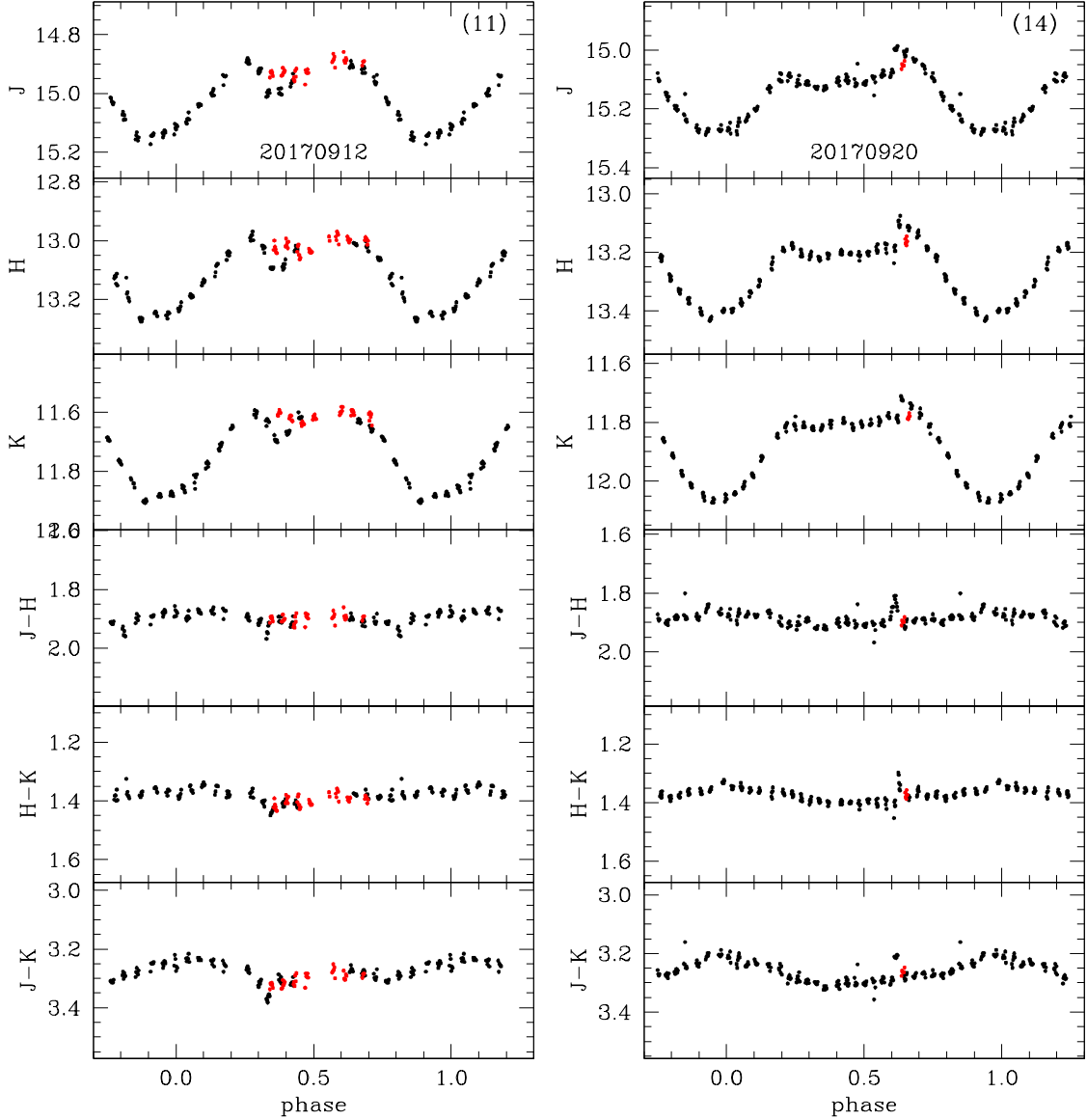}
\caption{Same as in Fig.~\ref{lc_jk_panel} for the nights with {\em JHK} data.}
\label{lc_jhk_panel}
\end{figure*}

\begin{deluxetable}{crcc}
 \tablecaption{Observations log.\label{table1}}
 \tablehead{
   \colhead{Date} &
   \colhead{Bands} &
   \colhead{Length of} &
   \colhead{Number of} \\
   \colhead{(YYYYMMDD)} &
   \colhead{ } &
   \colhead{observation (days)} &
   \colhead{measurements\tablenotemark{a}}
 }

 \startdata
   20160613 & \em  J/K  & 0.25 & 275/263 \\
   20160715 & \em  J/K  & 0.21 & 196/178 \\
   20160717 & \em  J/K  & 0.09 & 107/99 \\
   20160718 & \em  J/K  & 0.03 & 42/40  \\
   20161103 & \em  J/K  & 0.11 & 139/130 \\
   20161106 & \em  J/K  & 0.20 & 229/215 \\
   20161107 & \em   J   & 0.04 & 56  \\
   20161108 & \em  J/K  & 0.11 & 147/144 \\
   20170501 & \em  J/K  & 0.19 & 227/228 \\
   20170911 & \em J/H/K & 0.17 & 83/83/88  \\
   20170912 & \em J/H/K & 0.27 & 148/152/146 \\
   20170913 & \em   J   & 0.25 & 447 \\
   20170915 & \em   J   & 0.26 & 514 \\
   20170920 & \em J/H/K & 0.21 & 180/178/178 \\
 \enddata

 \tablenotetext{a}{In the corresponding IR bands.}
  
 \end{deluxetable}

\begin{deluxetable}{lc}[t!]
 \tablecaption{Photometry in {\em J} band. \label{table2}}
 \tablehead{
   \colhead{MJD} &
   \colhead{var-c1}
 }

 \startdata
    57552.772406 & 2.963 \\
    57552.772866 & 2.978 \\
    57552.773323 & 2.983 \\
    57552.773780 & 2.917 \\
    57552.774238 & 2.926 \\
 \enddata

 \tablecomments{Table~\ref{table2} is published in its entirety in the machine-readable format. A portion is shown here for guidance regarding its form and content.}
 
\end{deluxetable}

\begin{deluxetable}{lc}[t!]
 \tablecaption{Photometry in {\em H} band. \label{table3}}
 \tablehead{
   \colhead{MJD} &
   \colhead{var-c1}
 }

 \startdata
    58007.812930 & 1.323 \\
    58007.813360 & 1.338 \\
    58007.813795 & 1.336 \\
    58007.814254 & 1.343 \\
    58007.814689 & 1.351 \\
 \enddata

 \tablecomments{Table~\ref{table3} is published in its entirety in the machine-readable format. A portion is shown here for guidance regarding its form and content.}
 
\end{deluxetable}

\begin{deluxetable}{lc}[t!]
 \tablecaption{Photometry in {\em K} band. \label{table4}}
 \tablehead{
   \colhead{MJD} &
   \colhead{var-c1}
 }

 \startdata
    57552.777295 & 0.407 \\
    57552.777504 & 0.410 \\
    57552.777718 & 0.395 \\
    57552.777932 & 0.413 \\
    57552.778146 & 0.391 \\
 \enddata

 \tablecomments{Table~\ref{table4} is published in its entirety in the machine-readable format. A portion is shown here for guidance regarding its form and content.}
 
\end{deluxetable}

The orbital phases in all figures shown below were computed by using the best-fit ephemeris (Model 5) from \cite{ant19}. The folded {\em J} light curves for all observational nights are shown in Fig.~\ref{lc_j_panel}. In Figs.~\ref{lc_jk_panel},~\ref{lc_jhk_panel}, the light curves and colors are shown for those nights where observations covered the entire orbital cycle and were performed in more than one band. As the measurements in various bands were taken at slightly different moments of time, very short flares or rapid flux changes may lead to artificial color changes around these moments. Examples of such artificial changes are seen e.g. in Panel 2 of Fig.~\ref{lc_jk_panel} around orbital phases 0.02 and 0.7. A similar effect can be observed if the observations lasted more than one orbital period and the light curve varied between the adjacent orbital cycles. Such changes were indeed observed during some nights, see e.g. Panel 11 in Fig.~\ref{lc_jhk_panel}. For this reason, we computed the colors {\em before} folding the light curves with the orbital period. For nights where the duration of the observations was longer than one orbital period, the data points that lie further from the start of observations than the orbital period are shown in red.

To compare our IR data with the X-ray light curve of Cyg~X-3, we used the same {\em RXTE} ASM and {\em MAXI} data as in our previous paper \citep{ant19}. {\em RXTE} ASM provides 97996 flux measurements in A, B, C bands (A=1.3-3\,keV, B=3-5\,keV, C=5-12\,keV) covering the period from January 1996 to December 2011. The {\em MAXI}~ SCAN data constitute 27501 measurements in 2-4\,keV, 4-10\,keV, 10-20\,keV bands, covering the period from August 2009 to March 2019. {\em MAXI} observes the source nearly every International Space Station orbit (about 90\,min, subject to ISS orbit precession), which makes it possible to check the synchronicity of IR and X-ray variability. Since the orbital period of \mbox{Cyg X-3} is $\sim$4.8\,hr, we constructed quasi-simultaneous {\em MAXI} 2-20\,keV light curves from intervals of 20-30\,days centered approximately around the dates of our IR observations. The X-ray observation moments were also corrected to the solar system barycenter.

\begin{figure}
\centering
\includegraphics[width=8cm]{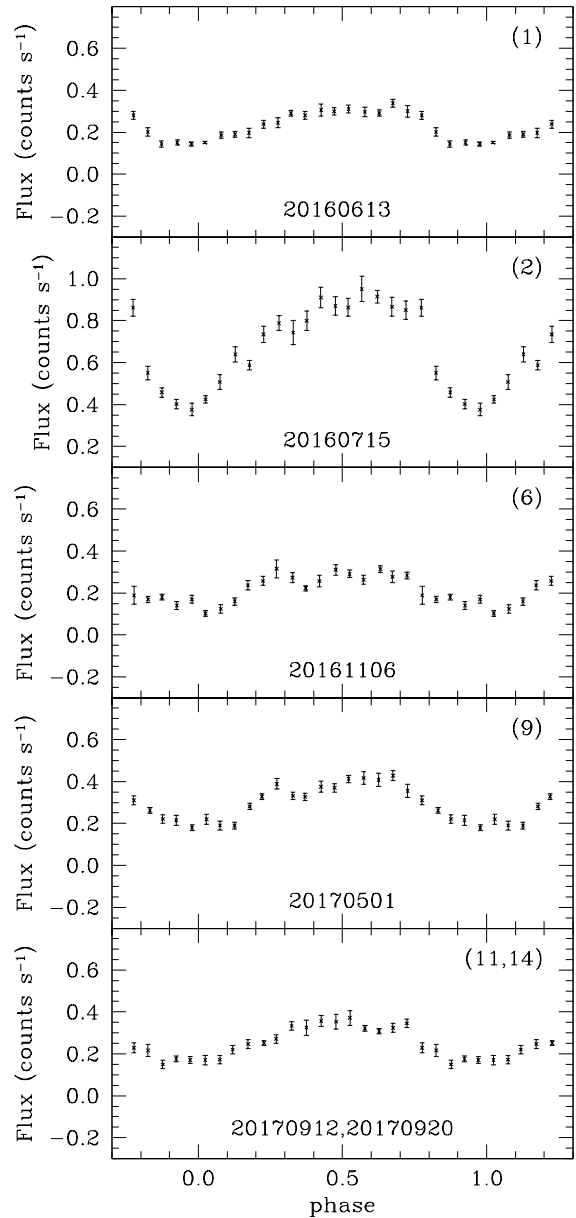}
\caption{{\em MAXI} 2-20\,keV light curves, quasi-simultaneous with the IR observations. The dates of the corresponding IR observations are shown in the bottom of every plot. The numbering is the same as in Figs.~\ref{lc_j_panel}-\ref{lc_jhk_panel}. For clarity, the mean light curves binned in 0.02 phase intervals are shown.}
\label{lc_MAXI_panel}
\end{figure}

The folded {\em MAXI} 2-20\,keV light curves, quasi-simultaneous with our IR ones, are shown in Fig.~\ref{lc_MAXI_panel}. To determine whether any phase shift exists between the IR and X-ray variations, we cross correlated our IR light curves with the {\em MAXI} ones. The average phase shift between the IR and {\em MAXI} light curves is $0.001 \pm 0.002$. Thus, the X-ray and IR orbital variability seems to be synchronous.

\section{The Mean X-ray and IR Light Curves}\label{sec:meanlc}

It is well known that the X-ray flux of Cyg~X-3 is subject to strong irregular variability in the form of flares and state changes. Yet at the same time the systematic orbital variability is remarkably well-defined (see, e.g., \citealt{zd12}). To check whether this is also true in the IR, and how the IR and X-ray light curves compare, we computed the mean normalized light curves using the {\em RXTE} ASM, {\em MAXI}, and our {\em JK} data ({\em H} data were not used since we have only two data sets which fully cover the orbital period). It should be noted that our IR observations were all made during the hard quiescent state of Cyg X-3. One night (2017 May 1st) was at the end of a giant flare occurred in April, when the system just returned to the hard spectral state \citep{trushkin17}. \cite{mccol10} reported that a change in state was accompanied by the near-IR color change such that $J-H$ was anti-correlated with $H-K_s$. Since our observations are all made in the hard state, these colors during the 3 nights with $JHK$ data are practically identical.

\begin{figure}
\centering
\plotone{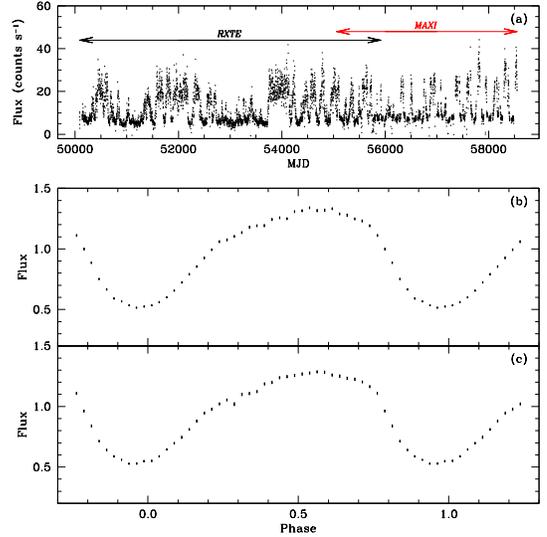}
\caption{{\em RXTE} ASM and {\em MAXI} light curves. (a) The combined light curve as function of Julian date. To decrease clutter, the daily averaged flux is shown. The energy bands are {\em RXTE} ASM 1.5-12\,keV and MAXI 2-10\,keV. (b) The mean folded normalized {\em RXTE} ASM light curve. (c) The mean folded normalized {\em MAXI} light curve.}
\label{lc_rxte_maxi}
\end{figure}

The energy bands and sensitivity of {\em RXTE} ASM and {\em MAXI} are different. Therefore, in order to compare their mean light curves, we chose the most similar energy ranges of 1.5-12\,keV and 2-10\,keV respectively. The {\em MAXI} flux was scaled so that its mean and standard deviation were equal to those of {\em RXTE} ASM flux. To mitigate the irregular flux changes, we followed the procedure similar to the one suggested by \cite{zd12}. First, all individual measurements were divided by a moving average with the window width of the moving average $\Delta t=2$\,days as recommended by \citeauthor{zd12}. Second, the time bin averages were computed with the time bin size equal to the phase bin size ($0.025$) of the intended folded light curve (time bin size = $0.025P$). Third, the time bin averages were combined in the phase bins, their mean and standard deviations computed. At every stage, the median filtration of the outliers was performed to avoid skewed mean values. The number of phase bins was chosen as a compromise such that every bin would contain a large number of individual measurements and yet the mean light curve would reproduce all essential details of the orbital variability. This number (40) is the same as the one used by \citeauthor{zd12}. The combined light curve as function of Julian date and the mean folded {\em RXTE} ASM and {\em MAXI}~ light curves are shown in Fig.~\ref{lc_rxte_maxi}. We will use this plot in the next section to discuss various aspects of the model we propose.

Note that when fitting the model to the X-ray data, we will use mean folded light curves in three energy bands. Since the {\em RXTE} ASM and {\em MAXI} bands do not match, we will use the {\em RXTE} ASM mean light curves in bands A, B, C. These light curves will be shown in Section \ref{sec:results}. Fitting the three {\em RXTE} ASM bands separately allows us (i)to check whether the parameters that should not depend on energy (such as the orbital inclination angle) are consistent in all bands; (ii)to see how the energy-dependent parameters (such as the optical depth of the WR wind) vary with the energy and whether these variations are consistent with theoretical expectations.

\begin{figure}
\centering
\plotone{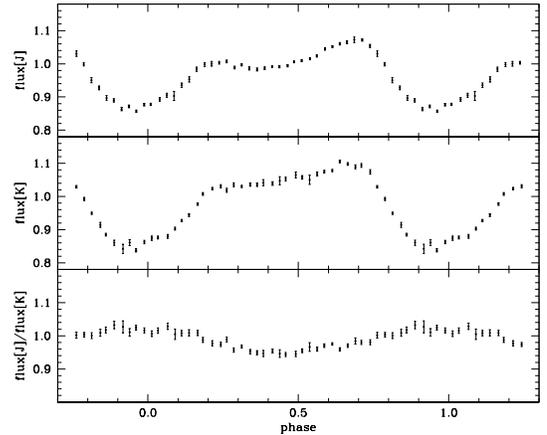}
\caption{The mean normalized {\em JK} light and color curves folded with the orbital period.}
\label{lccolobs}
\end{figure}

The mean normalized {\em JK} light curves were computed by using the data sets which covered the entire orbital period. The magnitudes were converted to arbitrarily scaled fluxes, the fluxes divided by the average flux, after that all data sets were combined and the mean light curves calculated. The mean normalized {\em JK} light curves are shown in Fig.~\ref{lccolobs}.

\section{The model}\label{sec:model}

\subsection{Qualitative considerations}\label{considerations}

As is obvious from Fig.~\ref{lc_rxte_maxi}, the mean normalized X-ray light curve is indeed remarkably stable over the course of 23 years. The common agreement is that the overall shape of the X-ray light curve can be explained in a simple model of scattering/absorption of the X-ray emission of an accreting relativistic companion by the spherically symmetric WR wind. A corresponding model was presented e.g., in \cite{zd12}. The broad minimum on the light curve near the zero orbital phase corresponds to the superior conjunction of the relativistic companion. However, scattering/absorption by a spherically symmetric wind produces a symmetric light curve and cannot explain the observed asymmetry of the ingress and egress of the minimum. 

Asymmetric X-ray light curves very similar to the one of Cyg~X-3 (fast ingress and slow egress) are also observed in two extra galactic WR+c candidates \mbox{NGC~300~X-1} \citep{carpano19} and \mbox{IC~10~X-1} \citep{steiner16}. Interestingly, a very similar variability was recently found in the first ULX WR+c candidate \mbox{CG~X-1} \citep{qiu19}. A notable difference with Cyg X-3 is that near the orbital phase zero (the X-ray source is behind the WR star), irregular dips occur where the X-ray flux drops to zero. \cite{qiu19} discuss various mechanisms leading to asymmetry of ingress and egress and to dips in X-ray binaries of different types. They conclude that the most plausible mechanism to explain the asymmetry of ingress and egress in case of WR+c binaries is absorption of the X-ray emission by the bow shock in front of the relativistic companion. The bow shock is caused by the fast orbital motion of the relativistic companion within the WR wind. In case of a super-Eddington accretion disk (which may be the case in \mbox{CG~X-1}) a strong disk wind can collide with the WR wind, leading to the formation of dense clumps along the line of centers. The zero X-ray flux in the dips is then explained by Compton-thick clumps which size is comparable to the size of the X-ray emitting region. In Cyg~X-3, the dips are not observed, probably because the accretion rate is much smaller and the disk wind is weak or absent (or accretion onto the relativistic companion occurs through the focused WR wind). However, it seems likely that the asymmetry of the X-ray minimum in Cyg X-3 is caused by the bow shock absorption.

Another feature of the X-ray light curve is a slight decrease of the flux near the orbital phase $\sim 0.4$ compared to what could be expected if wind-only scattering/absorption were present (see e.g. Fig.~4 in \cite{zd12}). \cite{vilhu13} suggested that the decrease could be due to absorption of the X-ray emission in the so called ``clumpy trail''. In their scenario, the relativistic jets trigger the formation of dense clumps in the WR wind which are then advected by the wind forming the absorbing trail. \cite{vilhu13}  numerically modelled the intersection of the line of sight with the trail during the orbit. They found that when using the orbital and jet parameters of Cyg X-3 known from previous studies, the ``clumpy trail'' manifests itself as an absorption feature centered at the orbital phase $\sim 0.4$, its full width is about half the orbital period.

Turning now to our mean {\em JK} light curves in Fig.\ref{lccolobs}, it is truly remarkable how qualitatively similar to the X-ray light curve they are. As in the X-ray domain, the fast ingress is followed by the slow egress, the absorption feature centered at the orbital phase $\sim 0.4$ is very clear.

\cite{vankerk93, vankerk96} suggested that the continuum IR flux and its variability in Cyg~X-3 can be explained by free-free emission of a two component spherically symmetric WR wind. Most of the wind is exposed to X-rays from the relativistic companion, this part is hot and highly ionized. The part of the WR wind that is in the shadow of the star is relatively cool. Near the orbital phase zero, the cold part of the wind is in front and partially obscures the hot part, which leads to a decrease of the free-free flux. 

This model, however, cannot fully explain our new IR light curves. First, it produces symmetric light curves. Second, at orbital phases near zero the cold part of the wind is most exposed to the observer and the binary should look more red ({\em J-K} should increase) than half a period later. Yet our {\em JK} light curves in Figs.~\ref{lc_jk_panel}, \ref{lc_jhk_panel}, \ref{lccolobs} show the opposite.

Thus, while the free-free emission of the WR wind in Cyg X-3 should be definitely present as part of the total IR flux, the similarity of the mean IR and X-ray light curves suggests that there exists another compact IR source in the binary, located near the relativistic companion and experiencing absorption in the same structures as the X-ray source. In this case, the observed IR color changes are naturally explained by the fact that the free-free absorption coefficient (corrected for stimulated emission) is proportional to the square of wavelength. 

\begin{figure}
\centering
\plotone{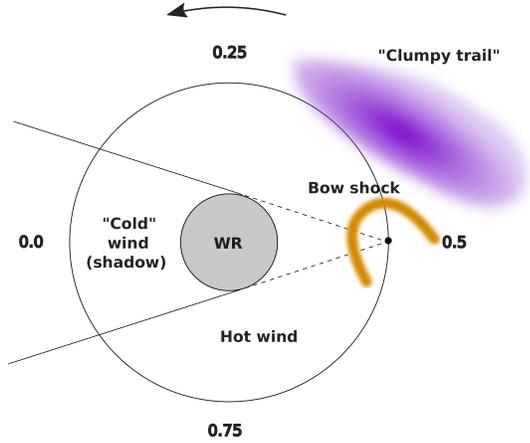}
\caption{Schematic view of Cyg~X-3 showing various emitting/absorbing components, projected onto the orbital plane. The relativistic companion is shown by the black dot. The orbit is shown by the black solid circle. The direction of the orbital motion is shown by the arrow. The numbers mark various orbital phases.}
\label{cartoon}
\end{figure}

We will postpone the discussion of a possible mechanism for the formation of a compact IR source until Section~\ref{sec:discussion}. Based on the above considerations, we take its existence as granted and will now qualitatively describe the models for calculating X-ray and IR fluxes, presented in the next subsections. A schematic view of the binary including various emitting and absorbing structures is shown in Fig.~\ref{cartoon}.

The X-ray model assumes that there is a point-like X-ray source located at the position of the relativistic companion. While the source is moving in orbit, the line of sight to the source can pass through the hot part of the WR wind, the bow shock, and the ``clumpy trail''. Variable absorption in these structures leads to variability of the emerging X-ray flux.

The IR model assumes that there are two sources of the IR emission: the spatially extent free-free emission of the hot and cold components of the WR wind and a point source located at the position of the relativistic companion. The flux of the latter can be absorbed by the same structures as in the X-ray model. 

Despite a significant progress in various aspects of theoretical studies of flows in single and binary stars achieved in last decade or so, there are still many uncertainties (for example, formation and evolution of clumps, unsteady wind-wind collision, accretion wakes etc.) that make a complete self-consistent model of systems like Cyg X-3 unfeasible. Therefore, we propose instead two simple models for calculating X-ray and IR fluxes, which, despite their simplicity, incorporate the qualitative ideas discussed above.

\subsection{X-ray}

The phase-dependent X-ray flux can be described by

\begin{equation}
 F^x(\phi_{obs}) = F^x_0e^{-\tau^x(\phi_{obs}-\phi_0)}\,,
 \label{eq1}
\end{equation}

where $F^x_0$ is the intrinsic X-ray flux of the compact source, $\tau^x(\phi)$ is the optical depth along the photon path, $\phi_{obs}$ is the observational orbital phase and $\phi_0$ is the phase of the superior conjunction of the relativistic companion. Recall that historically, a sine function was fit to observed X-ray light curves and the phase of its minimum was assigned as the orbital phase zero. Later on, the template light curve of \cite{klis89} replaced the sine method, but to maintain consistency, the template light curve was shifted so that using it to fit observed X-ray light curves would result in the same zero phase as the sine method. Thus, the observed phase zero does not necessarily correspond to the superior conjunction of the relativistic companion.

In our model, the X-ray emission of the compact source can be absorbed by the three structures: the WR wind, the bow shock, and the ``clumpy trail''. Accordingly,

\begin{equation}
 \tau^x(\phi) = \tau^x_w(\phi) + \tau^x_{bs}(\phi) + \tau^x_{ct}(\phi)
 \label{eq2}
\end{equation}

Assuming a circular orbit and a spherically symmetric wind, the WR wind absorption is 

\begin{equation}
 \tau^x_w(\phi) = \int\limits_a^\infty \frac{\alpha(r)rdr}{\sqrt{r^2-\xi^2}}\,,
 \label{eq3}
\end{equation}

in the phase interval 0.25-0.75 (the relativistic companion is in front of the sky plane passing through the WR center, $\phi=0$ at the superior conjunction of the relativistic companion) and

\begin{equation}
 \tau^x_w(\phi) = \int\limits_a^\infty \frac{\alpha(r)rdr}{\sqrt{r^2-\xi^2}} +2\int\limits_\xi^a \frac{\alpha(r)rdr}{\sqrt{r^2-\xi^2}}\,,
 \label{eq4}
\end{equation}

in the phase interval 0.75-1.25 (the relativistic companion is behind the plane). Here $a$ is the orbital separation, $\alpha(r)$ is the linear absorption coefficient, \mbox{$\xi=a\sqrt{1-\sin^2i\cos^2\phi}$} is the impact distance of the companion, $i$ is the orbital inclination. 

The main sources of the X-ray opacity are Compton scattering and bound-free absorption. Assuming that the ionization coefficient is constant throughout the WR wind,  $\alpha(r)$ is proportional to the wind density. By using the continuity equation, we obtain

\begin{equation}
 \alpha(r) = \sigma^x\frac{\dot{M}}{4\pi m_p\mu_er^2v(r)}\,, 
 \label{eq5}
\end{equation}

where $\sigma^x = \sigma_C+\sigma_{bf}$, $\dot{M}$ is the WR mass-loss rate, $m_p$ is the proton mass, $\mu_e$ is the mean electron molecular weight (as the wind composition is dominated by fully ionized Helium, $\mu_e\simeq 2$), and $v(r)$ is the velocity of the wind in the form of the commonly used $\beta$-law

\begin{equation}
 v(r) = V_\infty\left(1-\frac{R_*}{r}\right)^\beta\,,
 \label{eq6}
\end{equation}

where $V_\infty$ is the terminal velocity and $R_*$ is the radius of the WR star.

Substituting eq.\,(\ref{eq6}) to eq.\,(\ref{eq5}) and eq.\,(\ref{eq5}) to eqs.\,(\ref{eq3}), (\ref{eq4}), and expressing all distances in units of a, one can write (only equation (\ref{eq3}) is rewritten for shortness)

\begin{equation}
 \tau^x_w(\phi) = \tau^x_0\int\limits_1^\infty \frac{d(\frac{r}{a})}{(\frac{r}{a})(1-\frac{R_*}{r})^\beta\sqrt{(\frac{r}{a})^2-(\frac{\xi}{a})^2}}\,,
 \label{eq7}
\end{equation}

where

\begin{equation}
 \tau^x_0 = \sigma^x\left(\frac{\dot{M}}{4\pi m_p\mu_eV_\infty a^2}\right)a.
 \label{eq8}
\end{equation}

The term in brackets is the fiducial electron density of the WR wind (the density at a distance $a$ from the star under assumption that the wind velocity is $V_\infty$).

As mentioned in the previous subsection, self-consistent calculations of the bow shock and ``clumpy trail'' structures are unfeasible. Thus, to calculate the opacity of the bow shock and the ``clumpy trail'', we use simple approximations. Despite being phenomenological, they can still provide estimates of characteristic parameters such as optical depths and spatial scales of the corresponding structures.

We approximate the bow shock by a symmetric structure wrapped around the relativistic companion. The apex of the bow shock lies in the orbital plane. The direction to the apex is defined by the angle $\theta^x_{bs}$ between the line connecting the centers of the binary components and the line connecting the compact source and the apex. The angle can be roughly approximated by

\begin{equation}
 \tan\theta^x_{bs} = \frac{V_{orb}}{V_{wind}}\,,
 \label{eq:theta} 
\end{equation}

where $V_{orb}$ is the orbital velocity of the relativistic companion and $V_{wind}$ is the WR wind speed at the orbital separation. We approximate the optical depth of the bow shock by the bell-shaped raised cosine function

\begin{equation}
 \left\{
 \begin{array}{rclr}
  \tau^x_{bs}(\phi) & = & \tau^x_{bs}[ 1+\cos(\pi\frac{\psi(\phi)}{\psi^x_{bs}}) ]/2\,, & \psi(\phi)<\psi^x_{bs}\,, \\
  \tau^x_{bs}(\phi) & = & 0\,, & \psi(\phi)\geq\psi^x_{bs}\,,
 \end{array}
 \right.
 \label{eq:bowx}
\end{equation}

where $\tau^x_{bs}$ is the optical depth of the bow shock at its apex, $\psi(\phi)$ is the angle between the line of sight and the direction to the shock apex from the compact source, and $\psi^x_{bs}$ is the maximal opening half angle of the bow shock as seen from the position of the compact source. Thus, the bow shock optical depth gradually decreases from $\tau^x_{bs}$ to zero with increasing the angle between the direction to its apex and the line of sight.

\cite{vilhu13} showed that in their model the ``clumpy trail'' opacity leads to a decrease of X-ray flux around the orbital phase $\sim 0.4$ (see their Fig.~5 for the optical depth profile as function of the orbital phase). The authors made several model assumptions (e.g., that the WR wind velocity was constant and that the clumps created by the jets were fully advected by the WR wind etc.). Their results also depend on the adopted system parameters (the orientation of the jets and their opening angle etc.). Therefore, when approximating the ``clumpy trail'' absorption, we use the qualitative result of this work -- that is, the optical depth profile of the ``clumpy trail'' is a bell-shaped function of a certain width centered on a certain orbital phase. Similarly to the bow shock approximation, we use the raised cosine function

\begin{equation}
 \left\{
 \begin{array}{rcl}
  \tau^x_{ct}(\phi) & = & \tau^x_{ct}( 1+\cos(\pi\frac{\phi-\phi^x_{ct}}{\Delta\phi^x_{ct}}) )/2\,, \\
   & & \hspace{10mm} \phi-\phi^x_{ct} < \Delta\phi^x_{ct} \\
  \tau^x_{ct}(\phi) & = & 0\,, \hspace{6mm} \phi-\phi^x_{ct}\geq\Delta\phi^x_{ct}
 \end{array}
 \right.
 \label{eq:trailx}
\end{equation}

Here $\phi^x_{ct}$ is the orbital phase where the optical depth is maximal, $\Delta\phi^x_{ct}$ is its half width, $\tau^x_{ct}$ is the maximal optical depth of the ``clumpy trail'' at $\phi^x_{ct}$.

The model parameters are $\phi_0$, $i$, $R_*/a$, $\beta$, $\tau^x_0$, $\tau^x_{bs}$, $\psi^x_{bs}$, $\theta^x_{bs}$, $\tau^x_{ct}$, $\phi^x_{ct}$, $\Delta\phi^x_{ct}$, and a normalization parameter (see below).

\subsection{Infrared}

Following our qualitative considerations above, we compute the IR flux by

\begin{equation}
 F^{ir}(\phi) = F^{ff}_w(\phi) + F_{cs}(\phi)\,,
 \label{eq12}
\end{equation}

where $F^{ff}_w(\phi)$ is the free-free flux of the WR wind and $F_{cs}(\phi)$ is the IR flux of the compact source. The wind flux is

\begin{equation}
 F^{ff}_w(\phi) = \int\limits_0^{2\pi} d\eta \int\limits_0^\infty I(\phi,\eta,\xi)\xi d\xi\,,
 \label{eq13}
\end{equation}

where $\eta$ is the azimuth angle, $\xi$ is the impact distance of the photon path,

\begin{equation}
 I(\phi,\eta,\xi) = \int\limits_{l_1}^{\infty} j_\lambda(l,\xi,T)e^{-\tau^{ff}(l,\xi)}dl\,.
 \label{eq14}
\end{equation}

Here $j_\lambda(l,\xi,T)$ is the free-free emission of a unit volume within a unit solid angle per second per unit wavelength \citep{allen73}. If the impact distance $\xi > R_*$, $l_1=-\infty$. If  $\xi \leq R_*$ $l_1 = \sqrt{R^2_*-\xi^2}$. Depending on $\phi$, $\eta$, $\xi$, the photon path may lie completely in the hot part of the wind, within the shadow cone, or intersect the cone. For simplicity, we assume that the temperatures $T_h$ and $T_c$ of the hot and cold parts of the wind are constant throughout the respective volumes. Then, for a given segment of the photon path lying completely in the hot or cold part, using the Kirghoff law $j_\lambda=\chi_\lambda B_\lambda (T)$, one can write

\begin{equation}
  \begin{array}{rcl}
   I(\phi,\eta,\xi) & = & \int\limits_{l_1}^{l_2} \chi_\lambda(l,\xi,T) B_\lambda (T) e^{-\tau^{ff}(l,\xi)}dl\quad =\\
   & = & B_\lambda (T)(e^{-\tau^{ff}(l_2)}-e^{-\tau^{ff}(l_1)})\,,
  \end{array}
 \label{eq15}
\end{equation}

where $l_1$ and $l_2$ are the coordinates (along the photon path) of the end points of a segment. Substituting eq.\,(\ref{eq15}) to eq.\,(\ref{eq13}), accounting for possible intersections of the photon path with the shadow cone and expressing all distances in units of $a$, we obtain

\begin{equation}
 \begin{array}{l}
  F^{ff}_w(\phi) = a^2 B_\lambda(T_h) F_n(\phi)\,, \\
  F_n(\phi) = \int\limits_0^{2\pi} d\eta \int\limits_0^\infty \sum\limits_{i=1}^n\frac{B_\lambda(T_i)}{B_\lambda(T_h)}(e^{-\tau^{ff}_{i+1}}-e^{-\tau^{ff}_i}) \frac{\xi}{a} d(\frac{\xi}{a})\,,
 \end{array}
 \label{eq16}
\end{equation}

where $B_\lambda(T)$ is the Plank function, $T_i$ is the temperature of a photon path segment, $n$ is the number of segments ($n$ can vary from 1 to 3), $\tau^{ff}_i$, $\tau^{ff}_{i+1}$ are the optical depths at the end points of a segment. To compute the optical depths at the end points, one has to compute the optical depth of every segment starting from the one closest to the observer and then accumulate $\tau^{ff}_i$ by adding new segments. The optical depth of a segment is

\begin{equation}
 \Delta\tau^{ff}(r_1, r_2) = \int\limits_{r_1}^{r_2} \frac{\chi(\lambda, T)rdr}{\sqrt{r^2-\xi^2}}\,,
 \label{eq17}
\end{equation}

where $r_1$, $r_2$ are radius vectors of the segment's end points. In case the two end points are located in different hemispheres behind and in front of the sky plane passing through the center of the WR star and perpendicular to the line of sight, (\ref{eq17}) becomes slightly more complicated:

$$
 \Delta\tau^{ff}(r_1, r_2) = \int\limits_{r_0}^{r_1}\frac{\chi(\lambda, T)rdr}{\sqrt{r^2-\xi^2}}+\int\limits_{r_0}^{r_2}\frac{\chi(\lambda, T)rdr}{\sqrt{r^2-\xi^2}}\,,
$$

where $r_0$ is the radius-vector of the intersection point of the line of sight and the sky plane.

Note that $\Delta\tau^{ff}$ depends also on $\phi,\eta,\xi$; we omit these for clarity of notations. We also omit $r$ in the $\chi$ argument list for the same reason. The linear absorption coefficient $\chi(\lambda, T)$ in the IR domain (accounting for stimulated emission \citep{allen73}) is

\begin{equation}
 \chi(\lambda, T) \simeq 1.98\times 10^{-23} Z^2g\lambda^2T^{-3/2}n_en_i\,,
 \label{eq18}
\end{equation}

Given the overall simplicity of our model, we set the Gaunt factor $g=1$ and assume that the WR wind consists of (fully ionized) Helium only so $Z=2$, $\mu_e=2$, $n_i=n_e/\mu_e$. Using the mass continuity equation, expressing all distances in units of $a$ and substituting eq.\,(\ref{eq18}) to eq.\,(\ref{eq17}), we obtain

\begin{equation}
 \Delta\tau^{ff}(r_1, r_2) = b^{ff}_0\frac{\lambda_{\mu}^2}{T_4^{3/2}} \int\limits_{\frac{r_1}{a}}^{\frac{r_2}{a}} \frac{d(\frac{r}{a})}{(\frac{r}{a})^3(1-\frac{R_*}{r})^{2\beta}\sqrt{(\frac{r}{a})^2-(\frac{\xi}{a})^2}}\,,
\label{eq19}
\end{equation}

where $\lambda_{\mu}$ is the wavelength in microns, $T_4$ is the temperature in units of $10^4$\,K,

\begin{equation}
 b^{ff}_0 = 1.98\times 10^{-37}Z^2g\mu_e^{-1}\left(\frac{\dot{M}}{4\pi m_p\mu_e V_\infty a^2}\right)^2 a.
 \label{eq20}
\end{equation}

We introduce $b^{ff}_0$ in the form of (\ref{eq20}) so that this value does not depend on wavelength or temperature and instead is basically a measure of the WR mass-loss rate. As before, the term in brackets is the fiducial electron density.

When computing the absorbed IR flux of the compact source $F_{cs}(\phi)$, we use the formulae identical to (\ref{eq1}), (\ref{eq2}), except that the upper index ``$x$'' is replaced by ``$ir$''. The optical depth of the WR wind is computed similarly to the X-ray model (eqs. (\ref{eq3}), (\ref{eq4})) except that the integrals are replaced by the one in eq.\,(\ref{eq19}). When computing absorption by the bow shock and the ``clumpy trail'' we use the same approximations (\ref{eq:bowx}), (\ref{eq:trailx}) as in the X-ray model. We do not know what fraction of the total IR flux originates from the compact IR source. Thus we rewrite eq.\,(\ref{eq12}) in the form

\begin{equation}
 F^{ir}(\phi) = a^2 B_\lambda(T_h)(F_n(\phi) + f_0e^{-\tau^{ir}(\phi)})\,,
 \label{eq21}
\end{equation}

where

$$
 f_0 = \frac{F^{ir}_0}{a^2 B_\lambda(T_h)}.
$$

The model parameters are $\phi_0$, $i$, $R_*/a$, $\beta$, $f_0$, $T_h$, $T_c$, $b^{ff}_0$, $\tau^{ir}_{bs}$, $\theta^{ir}_{bs}$, $\psi^{ir}_{bs}$, $\tau^{ir}_{ct}$, $\phi^{ir}_{ct}$, $\Delta\phi^{ir}_{ct}$, and the normalization parameter (see below).

\subsection{The fitting method}

Before proceeding to the description of the fitting method, we will make some comments about the numerical aspects of the model. First, all integrals in our models were computed numerically by using an adaptive integration routine from GNU Scientific Library\footnote{https://www.gnu.org/software/gsl/.} allowing for integrands with integrable singularities. Second, the input data for both X-ray and IR models are normalized light curves which provide only relative change of the observed fluxes. Thus, in our models we cannot determine the absolute X-ray or IR flux. Instead, we compute normalized model light curves as follows:

\begin{enumerate}

\item Compute the X-ray model light curve at the phases of the observed one by using eq.\,(\ref{eq1}), with \mbox{$F^x_0=1$}. Likewise, compute the IR model light curve by using only the term in brackets of eq.\,(\ref{eq21}).

\item Compute the mean flux of a model light curve and divide the light curve by the mean flux.

\item Multiply the model light curve by the normalization parameter (which is treated as a free parameter of the model). The parameter is needed because the observed mean light curves are normalized by their weighted means. A model light curve is normalized by its unweighted mean since, by definition, the are no errors of the model fluxes. This difference can introduce a vertical shift between the normalized data and the model. Since the errors of the mean observed data points are similar, the normalization parameter is expected to be close to unity.

\end{enumerate}

Third, both X-ray and IR models have a large number of parameters. Fitting the models to the data in multidimensional parameter space may result in a solution falling in one of many local minima of the $\chi^2$ surface. Some parameter degeneracy is also possible. To cope with this problem we (i)implemented penalty functions, (ii)fixed some parameters, and (iii)used a two-step search for the global minimum of $\chi^2$ for every input light curve.

Some of the model parameters have natural limits. For example, the orbital inclination is limited to the range of $0-90^\circ$, the half-opening angle of the bow shock to the $0^\circ-180^\circ$ interval. Also, both our approximations of the bow shock and ``clumpy trail'' increase the optical depth in relatively narrow phase intervals. Without restrictions, the minimization routine can easily choose physically unrealistic parameters so the two intervals swap. To avoid these undesirable effects, we implemented the penalty functions which imposed limits on some parameters. The orbital inclination and the half-opening angle of the bow shock have been limited to the intervals given above. We limited the central phase of the ``clumpy trail'' profile to $0.3-0.5$ and its half width to $0-0.4$.

Some parameters are almost impossible to determine from the analysis of the light curves of \mbox{Cyg X-3}. For example, since there are no geometric eclipses in the system, the relative WR radius $R_*/a$ and the index $\beta$ in the velocity law influence the model only through the velocity law. However, some previous studies (e.g., \citealt{vankerk93, vankerk96} and \citealt{vilhu13}) used the assumption of constant $V(r)$ and could more or less satisfactorily fit the observed light curves. Thus, following \cite{zd12}, we assumed the fixed values of $R_* = 1.0\times 10^{11}$\,cm, $a = 3.0\times 10^{11}$\,cm (so $R_*/a = 1/3$) and $\beta=2.0$ (see the above paper for justifications). When solving the IR light curves, we also used fixed values for the temperatures of the hot and cold parts of the WR wind $T_h=10^5$\,K and $T_c=2\times10^4$\,K. IR model light curves are mainly influenced by their ratio (see eq.\,(\ref{eq16}), where the ratio of Plank functions in the IR is roughly equal to the temperature ratio). The problem is that, as shown by \cite{vankerk93, vankerk96}, a good light-curve solution in the WR wind free-free emission model can be found over a wide range of temperature ratios. The assumed values seem to be reasonable for the hot ionized and undisturbed (cold) parts of the WR wind \citep{ant04, kallman19}. We will discuss a potential influence of a change in our fixed parameters on the results in Section \ref{sec:caveats}.

To further restrict possible values of the model parameters, the mean X-ray and IR light curves were fitted in two stages: we started by fitting the mean {\em RXTE} ASM A, B, C light curves independently of each other. The orbital inclination $i$ and the phase of the superior conjunction of the relativistic companion $\phi_0$ should be independent of a band. Therefore, we proceeded to the second stage by fitting the A, B, C light curves at fixed $i$ and $\phi_0$, equal to the mean values found in the previous stage. {\em JK} light curves were also fitted at these fixed values of $i$ and $\phi_0$.

While the above measures allowed us to decrease the number of variable parameters and their allowable ranges, the remaining number of free parameters is still rather large: nine in the X-ray model at the initial independent fitting in the A, B, C bands and seven at the final fitting, and eight free parameters in the IR model. Thus, to increase our chances of finding the global $\chi^2$ minimum when fitting every input light curve, we implemented a two-step minimization.

First, the minimum was searched for by using the genetic algorithm \citep{char95}. We used an open source version of the algorithm presented by \cite{mohammadi17}. As usual with various flavours of the Monte Carlo method, there is no guarantee that the minimum found is indeed global. Some confidence in the results may be obtained if several runs of GA are performed. In our modeling, we run the GA three times on every input light curve with population sizes of $5\times10^5$, $7\times10^5$, $1\times10^6$. The results of the two latter were sufficiently close to provide some confidence that the global minimum was indeed found.

Second, we used the parameter values obtained with the GA, as a starting point for the Markov chain Monte Carlo (MCMC) method \citep{metropolis53, hastings70}. The method has the advantage that it allows one to get reliable posterior probability distributions of the model parameters in a Bayesian framework and is well suited for multi-dimensional problems. Of a crucial importance for the method is tuning the Markov chain parameters such as the proposed variance. In this study, we used the so called ``Adaptive Metropolis-Within-Gibbs'' \citep{mcmc11} variant of the MCMC algorithm which has the advantage that the proposed variance of every parameter is changing adaptively almost without user intervention. The algorithm varies only one model parameter at a time. Thus, to give a chance to every parameter to change between the saved chain states, we save the states every 100th iteration. Every MCMC run consisted of $10^5$ states which amounted to $10^7$ model evaluations. The burn-in period for various MCMC runs was usually between $10^4$ and $2\times 10^4$ states. The convergence of the chain in every run was verified by using the Gelman-Rubin statistics \citep{gelman92}. 

\section{Results}\label{sec:results}

\begin{figure}
\centering
\plotone{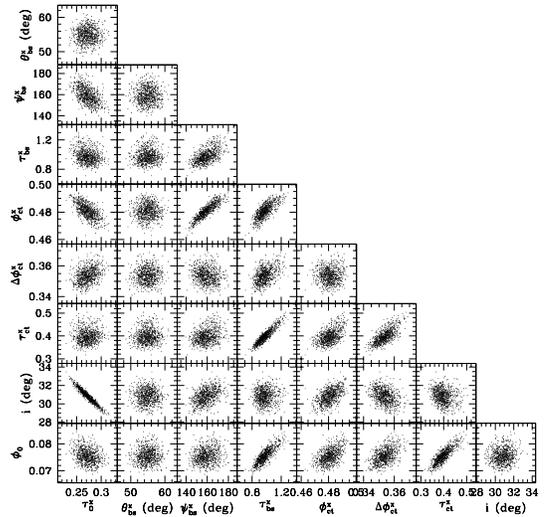}
\caption{A corner plot showing posterior probability distributions for the major X-ray model parameters in the {\em RXTE} ASM band C. Only 1/100 of MCMC states are shown to avoid cluttering. Correlation between some parameters is clearly visible.}
\label{xray_qq_plot}
\end{figure}

\begin{figure}
\centering
\plotone{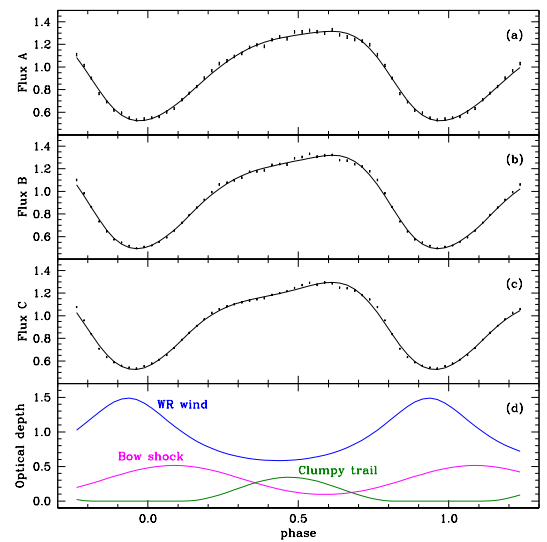}
\caption{The results of model fits at the second stage of minimization (see Table~\ref{tab_xsol}). (a)-(c): the mean normalized observed light curves and model fits (black solid lines) in A, B, C {\em RXTE} ASM bands. (d): the optical depths of the WR wind, the bow shock, and the ``clumpy trail'' for the best-fit model in the C band.}
\label{xray_sol}
\end{figure}

The results of the model fits are shown in Figs.~\ref{xray_qq_plot}, \ref{xray_sol}, Table~\ref{tab_xsol} (X-ray model), and in Fig.~\ref{ir_sol}, Table~\ref{tab_irsol} (IR model).

\begin{deluxetable}{lrrr}
 \tablecaption{Parameters of the X-ray model.\label{tab_xsol}}
 \tablehead{
   \multicolumn{4}{c}{Assumed parameters}
 }

 \startdata
   Par.            & \multicolumn{3}{c}{Value} \\ \hline
   $a$             & \multicolumn{3}{c}{$3.0\times10^{11}$ cm} \\
   $R_*$           & \multicolumn{3}{c}{$1.0\times10^{11}$ cm} \\
   $V_\infty$      & \multicolumn{3}{c}{$1700$ km\,s$^{-1}$} \\
   $\mu_e$         & \multicolumn{3}{c}{$2.0$} \\
   $\beta$         & \multicolumn{3}{c}{$2.0$} \\
   \hline
   \multicolumn{4}{c}{First stage: $i$ and $\phi_0$ as free parameters.} \\
   \hline
   \multicolumn{4}{c}{Fitted parameters} \\
   \hline
   Par.                  &   Band A    &   Band B    &  Band C \\ \hline
   $i$                   & $27.0^\circ \pm 2.0^\circ$ & $27.8^\circ \pm 1.0^\circ$ & $30.9^\circ \pm 0.8^\circ$ \\
   $\phi_0$              & $-0.055 \pm 0.006$         & $-0.059 \pm 0.003$         & $-0.076 \pm 0.003$ \\
   $\tau^x_0$            & $0.36 \pm 0.06$            & $0.37 \pm 0.03$            & $0.27 \pm 0.02$ \\
   $\tau^x_{bs}$         & $0.99 \pm 0.15$            & $0.94 \pm 0.08$            & $0.98 \pm 0.09$ \\
   $\theta^x_{bs}$       & $55 \pm 2^\circ$           & $56^\circ \pm 2^\circ$     & $55^\circ \pm 2^\circ$ \\
   $\psi^x_{bs}$         & $128^\circ \pm 8^\circ$    & $137^\circ \pm 7^\circ$    & $159^\circ \pm 7^\circ$ \\
   $\tau^x_{ct}$         & $0.33 \pm 0.05$            & $0.38 \pm 0.03$            & $0.40 \pm 0.03$ \\
   $\phi^x_{ct}$         & $0.48 \pm 0.01$            & $0.48 \pm 0.01$            & $0.48 \pm 0.01$ \\
   $\Delta\phi^x_{ct}$   & $0.35 \pm 0.01$            & $0.36 \pm 0.01$            & $0.35 \pm 0.01$ \\
   norm\tablenotemark{a} & $0.992 \pm 0.004$          & $0.973 \pm 0.002$          & $0.996 \pm 0.002$ \\
   $\chi^2$/dof          & $133/30$                   & $326/30$                   &  $454/30$    \\
   \hline
   \multicolumn{4}{c}{Second stage: fixed $i$, $\phi_0$.} \\
   \hline
     Par.            &  \multicolumn{3}{c}{Value} \\
    $i$              & \multicolumn{3}{c}{$29.5^\circ \pm 1.2^\circ$} \\
    $\phi_0$         & \multicolumn{3}{c}{$-0.066 \pm 0.006 $} \\
  \hline
   \multicolumn{4}{c}{Fitted parameters} \\
   \hline
   Par.                  &   Band A    &   Band B    &  Band C \\ \hline
   $\tau^x_0$            & $0.273 \pm 0.007$          & $0.318 \pm 0.004$          & $0.309 \pm 0.003$ \\
   $\tau^x_{bs}$         & $1.11 \pm 0.04$            & $1.08 \pm 0.04$            & $0.77 \pm 0.04$ \\
   $\theta^x_{bs}$       & $55 \pm 2^\circ$           & $56^\circ \pm 2^\circ$     & $55^\circ \pm 2^\circ$ \\
   $\psi^x_{bs}$         & $133^\circ \pm 5^\circ$    & $144^\circ \pm 6^\circ$    & $155^\circ \pm 8^\circ$ \\
   $\tau^x_{ct}$         & $0.34 \pm 0.03$            & $0.41 \pm 0.02$            & $0.34 \pm 0.01$ \\
   $\phi^x_{ct}$         & $0.49 \pm 0.01$            & $0.49 \pm 0.01$            & $0.47 \pm 0.01$ \\
   $\Delta\phi^x_{ct}$   & $0.36 \pm 0.01$            & $0.36 \pm 0.01$            & $0.35 \pm 0.01$ \\
   norm\tablenotemark{a} & $0.990 \pm 0.003$          & $0.971 \pm 0.002$          & $0.997 \pm 0.002$ \\
   $\chi^2$/dof         & $147/32$                   & $341/32$                   &  $470/32$    \\
 \enddata

 \tablenotetext{a}{Normalization parameter} 

\end{deluxetable}

\subsection{X-ray}

In Fig.~\ref{xray_qq_plot} an example of posterior distributions of the X-ray model parameters is shown for one of the MCMC runs on the {\em RXTE} ASM C band mean light curve. This run was performed at the first stage of minimization (the orbital inclination and the phase of the superior conjunction of the relativistic companion were free model parameters). Correlations between some model parameters are evident from the plot, the reasons for them are also quite clear. For example, the larger is the optical depth of the WR wind $\tau^x_0$, the smaller is the orbital inclination $i$ required to maintain the same light curve amplitude. The values of the fitted model parameters and their uncertainties in Table~\ref{tab_xsol} are determined from the empirical posterior distributions of the corresponding model parameters.

\begin{figure}
\centering
\plotone{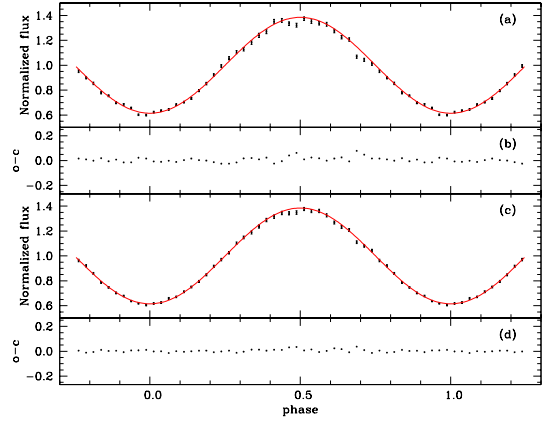}
\caption{The results of our numerical experiment. (a)the mean normalized light curve obtained from the simulated light curve (a sine function plus noise plus irregular variations) by applying the procedure of \cite{zd12}. The exact input light curve (a sine function) is shown by red solid line. (b)the deviations (o-c) of the mean filtered light curve from the exact one. (c)same as in panel (a) but for the simulated light curve not including irregular variations. (d)the deviations (o-c) for the light curves in Panel (c).}
\label{simulation_mav}
\end{figure}

The $\chi^2$ values of the best-fit model light curves in Table~\ref{tab_xsol} are quite large despite the good visual appearance of the model light curves in Fig.~\ref{xray_sol}. The reason for the large $\chi^2$ values might be underestimation of the uncertainty of the mean observed data points caused by filtering of irregular variability. The mean values of standard deviation of the mean data points are equal to $0.010$, $0.006$, and $0.005$ in A, B, C {\em RXTE} ASM bands respectively. The filtering was made by dividing the instant fluxes by the moving average. The latter is effectively a convolution and represents a low-pass frequency filter. Theoretically, its major drawback is that it allows to pass through a significant amount of the signal shorter than the width of the window. This can lead to the average light curve being not as smooth as expected. The situation is further aggravated by the fact that in our case the data are unevenly distributed over time. The width of the window is set in days; due to the uneven temporal distribution of data, the number of data points in each individual window varies.

To estimate how these effects can affect the accuracy of the mean normalized {\em RXTE} ASM light curves, we ran a numerical experiment as follows. First, we defined an artificial light curve as a sine function with the average value $1.0$ and the amplitude $0.4$ (similar to the amplitude of the actual observed orbital light curve) computed at all moments of observations. Second, we added Gaussian noise to every simulated data point, the standard deviation of the noise was defined by the relative error of the observed data (we used a single error $0.46$ for all simulated data points; this value is the average relative observational error in band A). Third, we multiplied the simulated light curve by the moving average values computed by using the actual data, i.e. added the observed irregular variations. Forth, we processed the simulated light curve with the same filtering/normalization method that was used for the original data.

The results of the simulation are shown in Panel (a) of Fig.~\ref{simulation_mav}. For comparison, Panel (c) shows the results of the same filtering procedure but applied to the simulated (noisy) light curve {\em before} adding the irregular variations. In both panels, the exact light curve (the simulated sine function without Gaussian noise) is also shown. The average standard deviations of both mean simulated light curves computed by the filtering procedure are almost equal ($0.010$ for the top light curve including the irregular variations and $0.009$ for the bottom light curve without irregular variations). However, the scatter of the light curve in the top panel around the exact light curve is clearly larger than that of the other. A better comparison of the scatters can be obtained by computing (o-c) values in both cases. For the light curve in the bottom panel, $\sigma\rm(o-c)=0.012$, for the light curve in the top panel $\sigma\rm (o-c)=0.022$. The former value is slightly larger than $\sigma=0.009$ obtained by the filtering procedure. This is because the procedure calculates the moving average and its value is not always equal to $1.0$ due to uneven temporal distribution of the data points. The latter value $\sigma\rm (o-c)=0.022$ exceeds the formal error produced by the filtering procedure ($0.010$) by more than two times. Note than in our experiment, the simulated noisy sine function did not include short flares which can make the underestimation of $\sigma$ even stronger.

To investigate the problems further, we conducted several additional experiments in which the window width of the moving average was changed from $0.5$ to $10$ days. In all cases, the scatter of the mean filtered simulated light curves exceeded the one computed by the procedure, by more than two times. Thus, we can safely suggest that the filtering procedure of \cite{zd12} underestimates the standard deviations of the mean filtered and normalized {\em RXTE} ASM light curves by at least two times (lower limit). It should be noted that despite this drawback the moving average filter is still the optimal filter for time domain encoded data if there exist no preferences to the weights of the individual data points within the moving average window \citep{dspguide}.

If $\sigma$ of all mean data points are proportionally increased by the factor of two, the $\chi^2$ values of the best-fit models must be decreased by four times. Applying this to e.g. the first stage results from Table~\ref{tab_xsol}, we obtain the null hypothesis probabilities in A, B, C bands $0.03$, $1.2\times 10^{-6}$, $1.2\times 10^{-11}$. This means that the first model can be accepted at the significance level 1\%, the other two are still rejected. This is not surprising given that some deviations of the model from the data are present in the phase interval $\sim 0.4-0.8$ in A, B, C bands, but more so in the two latter (Fig.~\ref{xray_sol}). It is hard to say whether the deviations are due to the overall simplicity of our model (unable to fully reproduce the small details of the regular variations) or due to underestimation of errors by the filtering procedure. Note however, that in the simulated filtered light curves the scatter in the same phase interval appears to be the largest.

The underestimation of the errors of the mean normalized light curves will not significantly affect the parameter values. This is because if $\sigma$ of all mean data points are increased by the same factor, the $\chi^2$ surface in the parameter space will be just scaled down and the position of its minimum (and hence the parameter values) will remain the same. The $\sigma$ values of different mean data points may change not proportionally. However, it is very unlikely that the disproportionality would be so large to significantly affect the best-fit parameter values. The parameter errors will increase. If our model were linear, the errors would increase by the same factor as the $\sigma$ values. In our non-linear model, the increase may be somewhat different. Albeit, obtaining more precise estimates of standard deviations of the mean data points is very hard or impossible. The current parameter errors are $1-\sigma$ errors obtained with the assumptions that the model is accepted and the errors of the mean data points are equal to the ones obtained with our filtering/normalizing procedure. We conclude that our parameter errors in Table~\ref{tab_xsol} can be underestimated by factor $\sim 2$.

Nevertheless, on the whole, our model was able to reproduce the characteristic features of the observed light curves even despite its overall simplicity. Let us first discuss the results obtained at the first stage of fitting (the three {\em RXTE} ASM bands fitted independently) and then the results of the second stage (the overall solution).

Our solutions at stage one can be considered as consistent only if the major geometrical parameters are consistent. These parameters are the orbital inclination angle, the phase of the superior conjunction of the relativistic companion, and the geometrical parameters of the bow shock and the ``clumpy trail''. Table~\ref{tab_xsol} shows that most of these parameters are indeed within the error bars of each other, except for the opening angle of the bow shock $\psi^x_{bs}$. Recalling that the parameter errors can be underestimated by about two times, the latter seem to be marginally consistent as well. Interestingly, the direction to the apex of the bow shock $\theta^x_{bs}$ in all three solutions is in very good agreement with the theoretical value $55.3^\circ$ calculated by eq.\,(\ref{eq:theta}) with the WR wind and orbital parameters used in our study (the terminal speed of the WR wind $V_\infty=1700$\,km$\,$s$^{-1}$ was taken from \citealt{zd12}).

The $\tau^x_0$ parameter characterising the optical depth of the WR wind, is maximal in band B. In the framework of our model, this is a consequence of the fact that the modulation amplitude (the ratio max/min) of the mean light curves is maximal in the same band (as also mentioned by \citealt{zd12}). The respective amplitudes are 2.49 (A), 2.67 (B), and 2.42 (C). Is this behavior of $\tau^x_0$ compatible with theoretical expectations? The X-ray bound-free opacity usually increases towards lower energies. \cite{zd12} explain the observational differences in the modulation amplitudes by a ``spectral component appearing in soft X-rays, which is probably re-emission in soft X-ray lines of the absorbed continuum by the stellar wind''. Another explanation could be that in a highly ionized wind, the bound-free absorption at low energies is much smaller than e.g. in the ISM and may even decrease towards low energies (see, e.g., \cite{ant04} and their Fig.~5). The behaviour of the optical depth of the bow shock $\tau^x_{bs}$ and the ``clumpy trail'' $\tau^x_{ct}$ is similar except that the former is nearly constant with the energy while the latter is monotonically increasing from band A to band C. This may mean that the temperatures of the bow shock and the ``clumpy trail'' are higher than that of the WR wind.

The consistency of the inclination angle and the phase of conjunction of the best-fit solutions in A, B, C bands allowed us to proceed to stage two and fit the three mean light curves using fixed values of these two parameters. The above discussion of the geometrical parameters of the bow shock and the ``clumpy trail'' and of the corresponding optical depths equally applies here; the general behaviour of these parameters is similar with some small quantitative differences.

The contribution of various wind structures to the total optical depth along the line of sight is shown in the lower Panel of Fig.~\ref{xray_sol} for the C band model. The maximal optical depth of the WR wind is reached at the phase of the superior conjunction of the relativistic companion ($\phi_0=-0.066$, see Table~\ref{tab_xsol}). The bow shock opacity decreases the total flux at the expected orbital phases, leading to an asymmetry of the ingress and egress. Note that at the inclination angle $\sim 30^\circ$ the line of sight passes peripheral areas of the bow shock. This is why the maximal optical depth of the bow shock in Fig.~\ref{xray_sol} is smaller than $\tau^x_{bs}$ in Table~\ref{tab_xsol} (recall that $\tau^x_{bs}$ is the optical depth at the bow shock apex).

The optical depth of the ``clumpy trail'' ($\sim 0.34-0.4$, see Table~\ref{tab_xsol}) is larger than the value $0.2$ obtained by \cite{vilhu13} for the high state of Cyg X-3 in the total ASM range. At the same time, their wind absorption ($1.82$) is larger than ours ($\sim 0.3$). This is because the model of \cite{vilhu13} did not account for the bow shock opacity. In our model, the opening angle of the bow shock is rather large, thus its opacity affects the light curve over a fairly wide phase interval, adding to the wind opacity and resulting in the final value comparable with that of \cite{vilhu13}.

From eq.\,(\ref{eq8}), by using $\mu_e$, $a$, the average $\tau^x_0=0.3$ from Table~\ref{tab_xsol}, and $V_\infty=1700$\,km$\,$s$^{-1}$, we obtain

$$
 \dot{M} \simeq 1.53\times 10^{-5} \sigma_T/(\sigma_C+\sigma_{bf})\,\rm M_\odot yr^{-1}.
$$

$\sigma_C+\sigma_{bf}$ is generally larger than $\sigma_T$ (the Thomson cross-section), thus the mass-loss rate is expected to be smaller than the numerical coefficient in this expression. Unfortunately, $\sigma_{bf}$ depends on the ionization structure of the wind which is hard to compute. 

\begin{figure}
\centering
\plotone{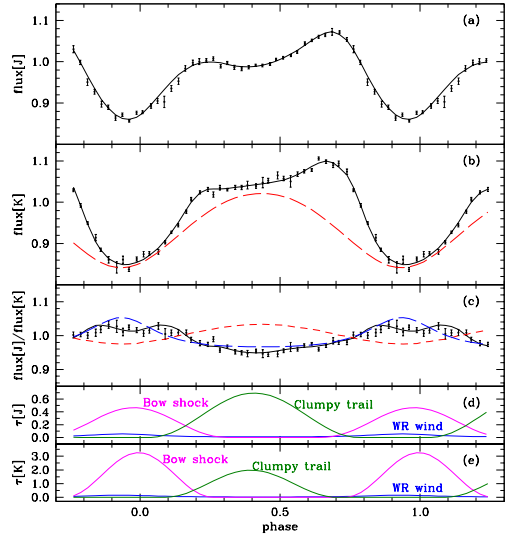}
\caption{The model fits of the mean normalized IR light curves. (a): {\em J} band; (b): {\em K} band. The red dashed line in Panel (b) is the component of the total model flux corresponding to the free-free emission of the WR wind component. We do not show the corresponding component in {\em J} band since it is located much lower than the observed and model light curves due to a larger fraction of the compact IR source flux (see Table~\ref{tab_irsol}); plotting it in scale would greatly increase the height of the figure. (c): {\em J-K} color of the data and the best-fit light curve (black solid line). The red dashed line shows the color of the free-free wind component, the blue long-dashed line shows the color of the compact IR source absorbed by the WR wind only (i.e. excluding absorption by the bow shock and the ``clumpy trail''. (d): the optical depths of the WR wind (blue line), of the bow shock (magenta line) and the ``clumpy trail'' (green line) in {\em J} band. (e): the same in {\em K} band.}
\label{ir_sol}
\end{figure}

\begin{deluxetable}{lrrr}
 \tablecaption{Parameters of the IR model. The assumed parameters are the same as in Table~\ref{tab_xsol}.\label{tab_irsol}}
 \tablehead{
   \multicolumn{4}{c}{Fitted parameters}
 }

 \startdata
   Par.                    &       {\em J}          &   {\em K}   &  \\ \hline
   $b^{ff}_0$              & $0.11\pm 0.03$         &  $0.10\pm 0.02$   \\
   $\tau^{ir}_{bs}$        & $1.2\pm 0.2$           &  $12\pm 2$   \\
   $\theta^{ir}_{bs}$, deg & $16^\circ \pm 3^\circ$ &  $22^\circ\pm 2^\circ$  \\
   $\psi^{ir}_{bs}$, deg   & $105\pm 3^\circ$       &  $93^\circ\pm 2^\circ$   \\
   $\tau^{ir}_{ct}$        & $0.7\pm 0.2$           &  $2.0\pm 0.3$   \\
   $\phi^{ir}_{ct}$        & $0.406\pm 0.004$       &  $0.395\pm 0.005$   \\
   $\Delta\phi^{ir}_{ct}$  & $0.37\pm 0.02$         &  $0.30\pm 0.02$    \\
   $f_0$                   & $0.7\pm 0.2$           &  $0.31\pm 0.08$  \\
   norm\tablenotemark{a}   & $1.000\pm 0.008$       &  $0.990\pm 0.002$ \\
   $\chi^2$/d.o.f          & $68.4/31$              &  $66.1/31$ \\
 \enddata

 \tablenotetext{a}{Normalization parameter}

\end{deluxetable}

\subsection{Infrared}

Several observations can be made from Fig.~\ref{ir_sol} and Table~\ref{tab_irsol} showing the best-fit IR model.

First, the ratios of the compact IR source flux to the WR wind flux in {\em J}, {\em K} bands (indicated by the $f_0$ values in Table~\ref{tab_irsol}) are different. Note that $f_0$ does not directly correspond to this ratio (see eq.\,(\ref{eq21})). Rather, the ratio is equal to the ratio of the second and first terms in eq.\,(\ref{eq21}). The mean values of the ratio in {\em J} band is 0.59 with the minimal and maximal values over the orbital cycle 0.54 and 0.66 respectively. In {\em K} band, the corresponding values are 0.16, 0.15, 0.17. This means that either the spectrum of the compact IR source is not that of free-free emission or that the temperatures of the compact source and of the hot part of the WR wind are different.

Second, despite the smaller visual ``depression'' of the observed light curve around the phase 0.4 in {\em K} band compared to {\em J} band, the optical depth of the ``clumpy trail'' in the former is larger than in the latter, in agreement with the expectation that the main absorbing mechanism in the ``clumpy trail'' is free-free absorption. Our simple approximation (\ref{eq:trailx}) does not assume any specific absorption mechanism. Mathematically, the optical depth in {\em K}~band is larger because the relative flux of the compact IR source in this band is smaller than in {\em J}~band, thus its absorbed fraction must be larger to produce the observed depression.

Third, the free-free optical depth is approximately proportional to the square of wavelength (see eq.\,(\ref{eq19})). Thus if the absorption mechanism of the bow shock and the ``clumpy trail'' is the free-free absorption, the corresponding optical depths in {\em K} band should be \mbox{$\sim (2.12/1.25)^2 = 2.9$} times larger than the optical depths in {\em J} band. While the ratio of the optical depths of the ``clumpy trail'' seems to be in agreement with this value, the ratio of the optical depths of the bow shock is not (see Table~\ref{tab_irsol}). Thus might be explained by the too rough approximation of the bow shock optical depth we used and/or by the fact that the free-free absorption is not the only absorption mechanism in the bow shock. Note that the angle of the direction to the bow shock in the IR $\theta^{ir}_{bs}$ is different from the theoretical expectation $55.3^\circ$. Still, the increase of the optical depth towards longer wavelengths is in agreement with theoretical expectations.

Fourth, the initial assumption in Section~\ref{considerations} that the color changes of the observed light curves are due to the free-free absorption of the compact IR source emission by the WR wind should be adjusted. In Panel (c) of Fig.~\ref{ir_sol} the colors of the free-free flux of the WR wind and the flux of the compact IR source absorbed by the wind are shown. As expected, the former becomes more red around the orbital phase zero, while the latter becomes more blue. However, the amplitude of these changes is such that they almost compensate each other. It is the absorption by the bow shock (increasing with wavelength) that makes the overall color of the model light curve more blue at these orbital phases.

Fifth, from Panel (c) of Fig.~\ref{ir_sol} the mathematical reason for the small angle $\theta^{ir}_{bs}$ defining the direction to the bow shock apex becomes evident. The component of the model light curve corresponding to the free-free absorption by the WR wind is unable to reproduce the steepness of the ingress of the observed light curve. Additional absorption is required at the orbital phases of the ingress, which results in the smaller values of $\theta^{ir}_{bs}$ compared to the X-ray solution.

Substituting $b^{ff}_0$ from Table~\ref{tab_irsol} to eq.\,(\ref{eq20}) we obtain an estimate of the WR mass-loss rate $\dot{M} = (0.96\pm 0.14)\times 10^{-5}\,\rm M_\odot yr^{-1}$.

We now compare this value with other existing estimates. \cite{szostek08} obtained $\dot{M} = (0.6-1.6)\times 10^{-5}\,\rm M_\odot yr^{-1}$ by modeling the X-ray spectra at different assumptions about the properties of the WR wind and the mass of the WR component. \cite{zd12} obtained the value $\dot{M} = (0.7-0.8)\times 10^{-5}\,\rm M_\odot yr^{-1}$ by modeling the orbital X-ray variability in the hard energy range, where the wind opacity is likely to be dominated by Thomson scattering. \cite{waltman96} obtained $\dot{M} \lesssim 1.0\times 10^{-5}\,\rm M_\odot yr^{-1}$ by analyzing the radio delays interpreted as being due to wind opacity. \cite{kol17} obtained $\dot{M} = (0.8-1.1)\times 10^{-5}\,\rm M_\odot yr^{-1}$ by fitting the IR spectrum of Cyg X-3 and assuming the distance to the system $d\simeq 7.4$\,kpc. These results are compatible with ours. On the other hand, \cite{ogley01} obtained $\dot{M}\simeq3.6\times 10^{-5}\,\rm M_\odot yr^{-1}$ by computing the $4\,\rm\mu m$ and $16\,\rm\mu m$ $ISO$ fluxes in a model of highly ionized WR wind. Their $\dot{M}$ for other WR wind models is even larger. However, their results depend on the adopted distance to Cyg X-3, and the authors assumed $d\simeq 10$\,kpc. In addition, a possible presence of the synchrotron emission of the jets (see below) can further decrease the calculated $\dot{M}$. \cite{vankerk93} obtained $\dot{M}\simeq 4.0\times 10^{-5}\,\rm M_\odot yr^{-1}$ by fitting the IR light curve of Cyg X-3 with a model which assumed that the IR flux variations are due to free-free emission of the WR wind consisting of two parts: the highly ionized wind exposed to the X-rays and the relatively cool wind in the shadow of the WR star. Our model is an extension of this model. However, our $\dot{M}$ is much lower. The reason is that the model of \citeauthor{vankerk93} does not include the point IR source present in our model. Thus, to fit the observed modulation amplitude, his model must use a larger wind optical depth (hence larger $\dot{M}$) compared to our model, where the flux variations due to the ``shadow wind'' provide only a fraction of the total modulation amplitude. 

Recall that neither of the IR models discussed above include wind clumping. Accounting for the clumping can reduce the model $\dot{M}$ even further (see below).

\section{Caveats of the model}\label{sec:caveats}

The first caveat is that the model used in the present study is semi-phenomenological. The approximations (\ref{eq:bowx}) and (\ref{eq:trailx}) are clearly very simple and can only serve as rough estimates of the actual absorption in the bow shock and ``clumpy trail''.

Another caveat is that the proximity of the binary components most likely makes the WR wind non-spherically symmetric and rather focused towards the relativistic companion. Indeed, the Bondi radius can be roughly estimated as

$$
R_B = \frac{2GM_x}{V^2_{wind}+V^2_{orb}}\,,
$$

where $M_x$ is the mass of the relativistic companion. Strictly speaking, this formula is valid for wide binaries where the Bondi radius is much smaller than the orbital separation. In a close binary $V_{orb}$ should be decreased by the rotational speed at the surface of the WR star (this would increase the Bondi radius). As we are interested in the lower estimate, we will use the formula as written above and also set $M_x=1.44\,\rm M_\odot$ (the minimal mass corresponding to the case of the relativistic companion being a neutron star). With our wind and orbital parameters, $R_B\simeq 0.3\,\rm R_\odot$. The actual radius is likely larger than this value. The flow of the WR wind material within the radius will be strongly affected by the gravitation of the relativistic companion resulting in the formation of either an accretion disk or a ``focused wind'' accretion onto the companion. 

Yet another caveat is the global asymmetry of the WR wind. \cite{vilhu21} have shown that the X-ray irradiation of the WR wind by the relativistic companion decreases the line force resulting in a decrease of the wind velocity towards the companion. 

Still, in the absence of a self-consistent model of the wind and various structures associated with the material flows in the system, our approach can hopefully improve our understanding of the processes occurring in the system.

Within the framework of our model there are two more circumstances that need to be discussed. First, we fit our IR model with the fixed temperatures of the hot and cold parts of the WR wind. The amplitude of the ``WR wind free-free emission'' component of the model light curve increases with the increase of the ratio of these temperatures. However, it seems unlikely that the temperature ratio can be much larger than the one used in the present study. Indeed, the amplitude of the ``WR wind'' light curve (the red dashed line in Panel (b) of Fig.~\ref{ir_sol}) is just barely smaller than the amplitude of the observed light curve. Increasing the temperature ratio would make the ``WR wind'' light curve amplitude larger than the observed one. The only way to reconcile the observed and model amplitudes would be to increase the IR flux of the compact IR source so the fraction of the WR wind flux would decrease. While it is hard to estimate the IR flux of the compact source quantitatively, it seems unlikely that it will be of the order or larger than the free-free flux of the WR wind.

Second, in our IR model we assumed that the WR wind is smooth with no clumps at all. As the free-free absorption/emission depends on the square of density, clumping may decrease the optical depth of the wind (and hence $b^{ff}_0$) resulting in a decrease of the mass-loss rate. We will return to this issue in the next section.

\section{Discussion}\label{sec:discussion}

What could be the origin of the compact IR source located near the relativistic companion? \cite{app97} suggested that in addition to the free-free emission of the WR wind, some IR emission can arise in the dense parts of the bow shock. In this scenario, short flares observed in IR can arise when the shock front encounters clumps of the WR wind. \cite{fender96} have suggested that IR flares can be a result of synchrotron emission from the relativistic electrons at the base of the jets. One could speculate that some more or less constant IR emission can also be created by this mechanism. Estimating the IR flux resulting from these mechanisms requires detailed hydrodynamic calculations which are beyond the scope of this paper.

Regarding the suggested bow shock, it is interesting to refer to \cite{zd10}. The authors explain the low cutoff energy in the X-ray spectra and X-ray power spectrum shape of Cyg X-3 by the existence of a small Thomson-thick plasma cloud surrounding the compact companion. The cloud is expected to be axially asymmetric with respect to the compact object. Given its small size (the estimated radius is $\sim 2\times 10^9\,\rm cm$) ``it cannot be the stellar wind itself. A likely origin of the cloud appears to be collision of the gravitationally focused stellar wind with the outer edge of an accretion disc.'' The resulting bulge cannot be a replacement of the suggested bow shock due to its small size. Also, the locations of the bow shock and the bulge are different. The bow shock is located more or less in the direction of the orbital motion of the compact companion while the bulge lags behind (see Fig. 7 of \cite{poutanen08} which \citeauthor{zd10} reference as an example of a similar structure in Cyg X-1). However, such structure will affect the X-ray orbital variability of the system and thus should possibly be included in future studies.

One of the most interesting and yet unresolved questions about Cyg X-3 is whether the relativistic companion is a neutron star or a black hole. \cite{zd13} estimated the mass of the companion to be $2.4^{+2.1}_{-1.1}\,\rm M_\odot$ based on the radial velocity curves of some X-ray \citep{vilhu09} and IR \citep{hanson00} spectral lines. However, later on, \cite{kol17} showed that the IR lines of \citeauthor{hanson00} did not reflect the motion of the WR component and were instead originating in the WR wind. Thus, the nature of the relativistic companion remained unclear.

We will now try to make some estimates based on the results of this and previous studies. We begin by recalling the well known formula relating the mass-loss rate $\dot{M}$ with the period change

\begin{equation}
 \frac{2\dot{M}}{M_{\rm WR}+M_{\rm C}} = \frac{\dot{P}}{P}
 \label{eq_massloss_pdot}
 \end{equation}

This assumes that the angular momentum of the WR star is removed by its stellar wind and that only a tiny fraction is accreted by the compact companion. \cite{ant19} refined the value of the period change $\dot{P}/{P}=1.02\times 10^{-6}\,\rm yr^{-1}$. By substituting this value to the above expression and using \mbox{$\dot{M} = 0.96\times 10^{-5}\,\rm M_\odot\, yr^{-1}$} from our current study, we obtain the total mass of the system $M_{\rm WR}+M_{\rm C}\simeq 18.8\,\rm M_\odot$.

The mass of the WR star can be estimated from the $\dot{M}-M$ relation for WR stars. \cite{nugis00} obtained this relation by fitting a sample of 34 WN stars (see their formula (23)). The mass-loss rates of individual stars were estimated by accounting for clumping. \cite{zd13} reanalyzed the same data and noted that the most massive WR stars in the sample strongly deviate from the $\dot{M}-M$ relation for the less massive stars. They redefined the relation by fitting a sub-sample that contained only WR stars with moderate masses $M_{\rm WR} < 22\rm M_\odot$, see their formula (6). The total mass of Cyg X-3 is also moderate, thus using the relation of \citeauthor{zd13} seems reasonable. This results in $M_{\rm WR}=(11.6\pm 1.2)\,\rm M_\odot$. Thus, the mass of the relativistic companion $M_{\rm C}\simeq 7.2\rm M_\odot$ pointing towards the black hole hypothesis. However, recall that our estimate of $\dot{M}$ was obtained with the assumption that the WR wind is smooth. If, due to wind clumping, the mass-loss rate is e.g. two times smaller, the same estimates would result in the total mass of the system \mbox{$M_{\rm WR}+M_{\rm C}= 9.4\,\rm M_\odot$} and \mbox{$M_{\rm WR}\simeq (9.2\pm 1.7)\,\rm M_\odot$}. This leaves too little room $M_{\rm C}\sim 0.2\,\rm M_\odot$ for even a neutron star. We can turn the question around and ask ourselves which $\dot{M}$ would be required to get the mass of the relativistic companion equal to the canonical mass of a neutron star $1.44\,\rm M_\odot$? It is easy to calculate that in this case the WR mass-loss rate should be equal $\dot{M}\simeq 0.57\times 10^{-5}\,\rm M_\odot\, yr^{-1}$, resulting in the total mass $\simeq 11.16\,\rm M_\odot$ and $M_{\rm WR}\simeq 9.74\,\rm M_\odot$. The WR mass is reasonable, thus we conclude that, unfortunately, considering the possibility of clumping, the above considerations are not capable of making a choice between the two hypotheses.

\cite{kol17} discussed the masses of the companions using the same approach but in more detail. In particular, they considered a case when a part of the WR angular momentum is accreted by the relativistic companion and then re-ejected in the form of accretion wind or jets. Then the formula (\ref{eq_massloss_pdot}) must be replaced by a more complex expression. In their Fig.~9 they plotted the regions of possible companion masses for different values of $\alpha$ -- the fraction of the mass lost directly from the system via the stellar wind. At reasonable values of $\alpha$ the limits for the masses of both components are very wide and do not really restrict the masses. They further tried to get some limits by restricting the value of $\beta=1-\alpha$ -- the fraction of mass accreted by the compact companion and re-ejected via accretion wind or jets. This fraction is a function of $\dot{M}$, $M_{\rm C}$, and the relative wind velocity at the location of the compact companion $v^2_{rel}=v^2_{orb}+v^2_{wind}$. Here $v_{orb}$ is the orbital velocity of the companion and $v_{wind}$ is the WR wind velocity at the location of the companion. By setting $\beta<0.25$ they produced the plots of the possible $M_{\rm WR}$, $M_{\rm C}$ at three values of $v_{rel}=1000$\,km\,s$^{-1}$, $800$\,km\,s$^{-1}$, and $700$\,km\,s$^{-1}$ in their Fig.~10. By using the semi-amplitude of the radial velocity curve defined by several X-ray spectral lines $K=400$\,km\,s$^{-1}$ they estimated the relative velocity $v_{rel}=750-1000\,\rm km\,s^{-1}$ corresponding to the the orbital inclination angles $i=30^\circ-70^\circ$. Then $M_{\rm C}\lesssim 10\,\rm M_\odot$ for $v_{rel}=1000\,\rm km\,s^{-1}$ and $M_{\rm C}\lesssim 5\,\rm M_\odot$ for $v_{rel}=750\,\rm km\,s^{-1}$ (see their Fig.~10). The orbital inclination angle in our best-fit models is close to $30^\circ$ corresponding to the former estimate.

\section{Conclusions}\label{sec:concl}

We have studied X-ray and near-IR light curves of Cyg X-3 using archive {\em RXTE} ASM and {\em MAXI} observations and {\em JHK} photometric data obtained with the 2.5m telescope of the Caucasian Mountain Observatory (CMO) of the Lomonosov Moscow State University Sternberg State Astronomical Institute. We have found that the mean IR light curves of the binary are well defined despite strong irregular variability observed in individual orbital cycles. This  is similar to the X-ray domain.

It is shown that the shape of the mean IR light curves is qualitatively similar to the shape of the X-ray ones. The shape of the X-ray light curves can be explained if, apart from scattering/absorption in the WR wind, the X-ray emission of the compact source is also attenuated by additional wind structures: the bow shock and the so-called ``clumpy trail''.  The similarity of the IR and X-ray light curves suggests that in addition to the free-free emission of the WR wind, another compact IR source is present in the system located near the relativistic companion. The flux of this IR source is then affected by the same wind structures as the flux of the X-ray compact source. The existence of a compact IR source is also confirmed by the color shift of the IR flux to blue near the orbital phase zero. It is important that this is a qualitative conclusion which does not depend on any specific modeling.

To analyse the X-ray and IR light curves, we developed a model which semi-phenomenologically accounts for the scattering/absorption of X-ray and IR fluxes in the WR wind and in two additional structures: the bow shock and the ``clumpy trail''. Despite its simplicity, the model is useful for determining the characteristic parameters of these structures. By analysing the mean X-ray and IR light curves we were able to find the orbital phase of the superior conjunction of the relativistic companion $\phi_0=-0.066\pm 0.006$, the orbital inclination angle $i=29.5^\circ\pm 1.2^\circ$ and the WR mass-loss rate \mbox{$\dot{M} = (0.96\pm 0.14)\times 10^{-5}\,\rm M_\odot\, yr^{-1}$}.

By using the relations between $\dot{M}$ and the rate of the period change and between $\dot{M}$ and the WR mass, we estimated the probable mass of the relativistic companion $M_{\rm C}\simeq 7.2\rm M_\odot$ which points towards the black hole hypothesis for the relativistic companion. However, this estimate is based on the assumption of a smooth WR wind. Considering the uncertainty associated with clumping, the mass-loss rate can be lower which leaves room for the neutron star hypothesis.

Clearly, our quantitative results depend on the adopted model. They can be considered as one of the steps towards using more advanced models that take into account the presence of a compact IR source and additional wind structures.

\begin{acknowledgments}

The authors are grateful to the anonymous referee for an insightful report, which helped us to significantly improve the paper. This research has made use of {\em RXTE} ASM data obtained through the High Energy Astrophysics Science Archive Research Center Online Service, provided by the NASA/Goddard Space Flight Center, and of the {\em MAXI} data provided by RIKEN, JAXA and the {\em MAXI} team. The work of IA (observations and modeling), ACH (modeling), and AT (observations and data reduction) was supported by the RSF grant 17-12-01241. The work of EA was supported by the Scientific Educational School of Lomonosov Moscow State University “Fundamental and applied Space Research”.

\end{acknowledgments}

\bibliography{cyg_x3_IR}

\end{document}